\documentclass[apj]{emulateapj}
\usepackage{apjfonts}

\newcommand{\beq}{\begin{equation}}
\newcommand{\eeq}{\end{equation}}
\newcommand{\beqn}{\begin{eqnarray}}
\newcommand{\eeqn}{\end{eqnarray}}
\newcommand{\pd}{\partial}
\newcommand{\bi}{{\bf I}}
\newcommand{\bx}{{\bf x}}
\newcommand{\bX}{{\bf X}}
\newcommand{\AU}{{\rm AU}}

\newcommand{\ovl}[1]{ {\overline{#1}} }

\newcommand{\bracket}[1]{\langle #1 \rangle}
\newcommand{\eqref}[1]{(\ref{#1})}

\newcommand{\dint}{\displaystyle\int}
\newcommand{\dfrac}[2]{ {\displaystyle\frac{#1}{#2}} }

\newcommand{\pfrac}[2]{ \left(\dfrac{#1}{#2}\right) }
\newcommand{\eps}{\epsilon}
\newcommand{\vareps}{\varepsilon}
\renewcommand{\bullet}{0}

\defcitealias{OST07}{OST07}
\shorttitle{COAGULATION AND POROSITY EVOLUTION OF DUST AGGREGATES}
\shortauthors{OKUZUMI ET AL.}
\slugcomment{Received 2009 July 2; accepted 2009 November 3}

\begin{document}
\title{Numerical modeling of the coagulation and porosity evolution of dust aggregates}
\author{Satoshi Okuzumi\altaffilmark{1}, Hidekazu Tanaka\altaffilmark{2}, and Masa-aki Sakagami\altaffilmark{1}}
\email{satoshi.okuzumi@ax2.ecs.kyoto-u.ac.jp}
\altaffiltext{1}{Graduate School of Human and Environmental Studies, Kyoto University,
Yoshida-nihonmatsu-cho, Sakyo-ku, Kyoto 606-8501, Japan}
\altaffiltext{2}{Institute of Low Temperature Science, Hokkaido University, Sapporo 060-0819, Japan}
\begin{abstract}
Porosity evolution of dust aggregates is crucial in understanding dust evolution 
in protoplanetary disks.
In this study, we present useful tools to study the coagulation and porosity evolution
of dust aggregates.
First, we present a new numerical method for simulating dust coagulation and porosity evolution
as an extension of the conventional Smoluchowski equation.
This method follows the evolution of the mean porosity for each aggregate mass
simultaneously with the evolution of the mass distribution function. 
This method reproduces the results of previous Monte Carlo simulations with much less 
computational expense. 
Second, we propose a new collision model for porous dust aggregates 
on the basis of our $N$-body experiments on aggregate collisions.
As the first step, we focus on ``hit-and-stick'' collisions, which involve
neither compression nor fragmentation of aggregates.
We first obtain empirical data on porosity changes between the 
classical limits of ballistic cluster--cluster and particle--cluster aggregation.
Using the data, we construct a recipe for the porosity change due to general hit-and-stick collisions
 as well as formulae for the aerodynamical and collisional cross sections.
Our collision model is thus more realistic than a previous model of Ormel et al.
based on the classical aggregation limits only.
Simple coagulation simulations using the extended Smoluchowski method 
show that our collision model explains the fractal dimensions 
of porous aggregates observed in a full $N$-body simulation and a laboratory experiment.
By contrast,  similar simulations using the collision model of Ormel et al. result in much less porous aggregates,
meaning that this model underestimates the porosity increase upon unequal-sized collisions.
Besides, we discover that aggregates at the high-mass end of the distribution 
can have a considerably small aerodynamical cross section per unit mass 
compared with aggregates of lower masses.
This occurs when aggregates drift under uniform acceleration (e.g., gravity) and their collision 
is induced by the difference in their terminal velocities.
We point out an important implication of this discovery for dust growth in protoplanetary disks.

\end{abstract}
\maketitle

\section{Introduction}
It is a widely accepted idea that the first step of planet formation in protoplanetary disks  
involves the collisional growth of submicron/micron dust grains into macroscopic aggregates.
A standard scenario is that dust grains coagulate by mutual sticking,
gradually settle toward the midplane of the disk, and form a dense dust layer (e.g., \citealt*{WC93}).
It is still an open issue whether subsequent planetesimal formation is achieved 
by the gravitational instability of the layer \citep{S69,GW73,S98,YS02} or
 by the direct collisional growth of the aggregates \citep{WC93,SV97,BDH08}. 
To address this issue, further understanding on earlier evolutionary stages is needed.
 
Most of the previous studies have simplified dust aggregates as compact, nonporous spheres.
However, both numerical and laboratory experiments have revealed
that aggregates are not at all compact, but have an open, 
fluffy structure (for a review, see \citealt*{Blum04,DBCW07}).
This is particularly true for aggregates formed at an early growth stage where 
the collisional velocity is so low that collisional compression is negligible.
It has been observed in $N$-body \citep{KPH99} as well as experimental
 \citep{WB98,Blum+98,Blum+00} studies that the outcome is an ensemble of
fractal aggregates with the fractal dimension of $D \la 2$.
This fractal growth lasts until the impact energy becomes high enough to 
cause collisional compaction \citep{SWT08}.
The porosity change due to compressive as well as destructive collisions are also studied
 by some authors \citep[e.g.,][]{Wada+07,Wada+08,Wada+09,SWT08,PD09} 
 using $N$-body simulations including monomer surface interactions  \citep{DT97}.

There are several strong reasons why the porosity evolution of aggregates is important 
in studying dust evolution in protoplanetary disks. These include the following.
\begin{enumerate}
\setlength{\itemsep}{-0pt}
\item
The porosity affects the aerodynamical property of aggregates.
In a gas disk, the motion of a small aggregate is controlled by 
the friction force from the ambient gas.
The friction coefficient can vary by many orders of magnitude
 depending on whether the aggregate is compact or fluffy.
For example, a fractal ($D\la2$) aggregate made of one thousand grains
receives a friction force an order of magnitude stronger than a compact aggregate of the same mass.
This difference is significant when one considers the formation of sedimentary dust layer on the disk midplane.
\item
The porosity even affects the turbulence of protoplanetary disks.
Magnetorotational instability (MRI; \citealt*{BW91}),
 which is recognized as the most likely mechanism driving disk turbulence,
operates only in a region where the ionization degree is sufficiently high \citep{Gammie96}. 
The ionization degree strongly depends on the surface area of dust aggregates, 
 since the recombination of ionized gases efficiently takes place on dust surfaces.
Again, a fractal ($D\la2$) aggregate made of one thousand grains
absorbs ionized gases an order magnitude more efficiently than a compact grain of the same mass.
For this reason, compact dust growth tends to make the gas environment turbulent \citep{Sano+00},
 while fractal dust growth tends to keep the environment laminar \citep{Okuzumi09}.
This difference is critical to the evolution of dust itself,
since the turbulence causes the diffusion and collisional fragmentation of aggregates \citep{BDH08,J+08}.
\item
The fractal nature of aggregates matters when we try to interpret
 observed emission spectra of protoplanetary disks.
For example, the $10\micron$ spectral feature characteristic of submicron silicate grains 
quickly vanishes in compact aggregation, 
but only slowly in fractal $(D\sim 2)$ aggregation \citep{Min+06}.
This means that the conventional compact dust model generally underestimates the degree of dust growth
in the observed objects.
\end{enumerate}
Thus, the consistent modeling of the growth and porosity evolution of dust aggregates
is needed to better understand and predict dust evolution in protoplanetary disks.

Theoretically, however, the consistent modeling is not an easy task.
First of all, we must consider the evolution of aggregates in the 
two-dimensional (2D) parameter space of mass and porosity. 
However, such 2D calculation is generally much more elaborate than one-dimensional
(1D; i.e., mass only) one.
Furthermore, such calculation requires a reliable {\it collision model} for porous aggregates.
Here, a ``collision model'' refers to a set of (1) the definition of the ``porosity,'' or ``volume,'' 
of an aggregate, (2) a recipe that determines the porosity change due to collisions,
and (3) formulae for collisional and aerodynamical cross sections that determine the collision 
rate of porous aggregate pairs.
To obtain a reliable collision model, one needs a sufficient amount of empirical data on aggregate collisions.

Because of the above theoretical difficulty,
there are only a few studies addressing the coagulation of porous dust aggregates in the literature.
A 2D simulation on porous dust growth was first done by
\citet{Ossenkopf93} in the context of dust extinction in molecular clouds.
He used a collision model based on the $N$-body simulations of
the ballistic cluster--cluster and particle--cluster aggregation (BCCA and BPCA; \citealt*{Meakin91}).
Ormel et al. (2007; hereafter, \citetalias{OST07}) have 
presented a Monte Carlo method for studying porosity evolution of dust aggregates in protoplanetary disks.
In this method, porous aggregates are represented by the same number of computational particles
with mass and porosity, and the collisions among the particles are successively 
calculated using random numbers.
The change in porosity on each individual collision is determined from a recipe 
based on previous $N$-body experiments on aggregate collisions \citep{Ossenkopf93,DT97}.
The Monte Carlo approach has been further developed independently by \citet{OS08} and \citet{ZD08} 
to achieve a high dynamical range in the parameter space.
The compression/fragmentation model of \citetalias{OST07} has been recently 
revised by \citet{OPDT09} on the basis of numerical experiments on low-mass aggregate 
collisions \citep{PD09}.  

In this study, we provide new theoretical tools useful
for studying the collisional evolution of porous dust aggregates.
First, we present a new numerical method to solve the growth and porosity evolution of aggregates.
This method is an extension of the conventional Smoluchowski method, a method 
commonly used for 1D simulations of dust aggregates \citep[e.g.,][]{THI05,DD05,BDH08}.
The core of our new method is the approximation 
that the porosity distribution is narrow for each aggregate mass.
With this approximation, we derive a closed set of moment equations of the 2D Smoluchowski equation
for the mass distribution function and the averaged mass-porosity relation.
These equations are formally similar to the conventional 1D Smoluchowski equation,
and makes it possible to adopt well-established algorithms for 1D problems. 
We confirm that this approach indeed reproduces the results of the Monte Carlo
simulations by \citetalias{OST07} with much less computational effort. 
Thus, our method will be particularly useful for adding the porosity evolution 
to global simulations including radial drift \citep[e.g.,][]{BDH08} or
 coupled simulations involving hydrodynamic calculation \citep[e.g.,][]{J+08}. 
Second, we provide a new collision model based on our $N$-body experiments for
various types of aggregate collisions.
As the first step of the modeling, we focus on ``hit-and-stick'' collision, i.e., 
low-velocity collision involving neither compression nor fragmentation.
In a typical protoplanetary disk, the hit-and-stick picture well represents
the growth of dust aggregates up to several centimeters \citep{SWT08}.
In contrast to previous studies, we first use empirical data on the porosity evolution 
between the BCCA and BPCA limits.
We will see that  the numerical simulations of porosity evolution using our collision model
well agree with the results of previous full $N$-body and laboratory experiments.
Extension of our model to compressive and destructive collisions will be made in the future work.

This paper is organized as follows.
In Section 2, we describe our extended Smoluchowski method and its numerical implementation.
In Section 3, we use this method to try to reproduce the results of the Monte Carlo simulations by
\citetalias{OST07}.
In Section 4, we present a new porosity model as well as the results of our $N$-body simulations
from which we construct the porosity increase recipe.
In Section 5, we compare this collision model with that of \citetalias{OST07}.
Section 6 is devoted to the summary. 
\section{The extended Smoluchowski method for porous dust growth}
In this section, we describe a new method to solve the coagulation and porosity evolution
of dust aggregates.
A comparison between our method and the Monte Carlo method 
is  made in Section 3.

\subsection{The Multi-dimensional Smoluchowski Equation}
We denote the internal structure of each aggregate by a set of parameters 
$\bi = \{ I^{(1)},I^{(2)},\dots \}$.
Conventional coagulation models adopted the mass $M$ of an aggregate as 
a unique parameter, i.e., $\bi = \{ M \}$.
This study treats the volume $V$ as an additional parameter for porous aggregates.
We assume that the internal structure of an aggregate created by a collision is 
uniquely determined by those of the collided ones.
This means that the structure parameter of the new aggregate,
 $\bi_{1+2}$ is a function of those of the old ones, $\bi_1$ and $\bi_2$.
 
We denote the distribution function of aggregates by $f(\bi)$.
The evolution of $f(\bi)$ is determined by the multidimensional Smoluchowski equation
\beqn
\frac{\pd f(\bi)}{\pd t} 
&=& \frac{1}{2}\int d\bi' d\bi''\; K(\bi';\bi'')f(\bi')f(\bi'') \nonumber \\
&&\qquad \times \delta[\bi_{1+2}(\bi';\bi'')-\bi]
\nonumber \\
&& - f(\bi)\int d\bi'\; K(\bi;\bi')f(\bi'),
\label{eq:coag0}
\eeqn
where $K(\bi_i;\bi_j)=K(\bi_j;\bi_i)$ is the collision kernel defined
as the product of the collisional cross section $\sigma_{\rm coll}$ and the relative speed $\Delta u$ 
of colliding pairs.
The first and second terms in Equation \eqref{eq:coag0} represent 
the ``gain'' and ``loss'' of aggregates with $\bi$ due to collisions, respectively.

If $\bi$ only contains $M$, Equation \eqref{eq:coag0} reduces to the conventional, 1D
 Smoluchowski equation
\beqn
\frac{\pd f(M)}{\pd t} 
&=& \frac{1}{2}\int_0^M dM'\; K(M';M-M')f(M')f(M-M') \nonumber \\
&& - f(M)\int_0^\infty dM'\; K(M;M')f(M'),
\label{eq:coag1}
\eeqn
where we have used the mass conservation $M_{1+2}(M',M'') = M' + M''$.
It is easy to check that Equation \eqref{eq:coag1} ensures the conservation of the
total mass density $\rho_d \equiv \int Mf(M) dM$.

Now we consider the 2D case, $\bi = (M,V)$.
In this case, equation \eqref{eq:coag0} is written as
\beqn
\frac{\pd f_{M,V}}{\pd t} 
&=& \frac{1}{2}\int_0^M dM' \int dV'dV''\; K^{M',V'}_{M-M',V''} f_{M',V'}f_{M-M',V''} 
\nonumber \\
&& \quad\times \delta[(V_{1+2})^{M',V'}_{M-M',V''}-V]
\nonumber \\
&& - f_{M,V}\int dM'dV'\; K^{M,V}_{M',V'}f_{M',V'},
\label{eq:coag2}
\eeqn
where $f_{M,V} \equiv f(M,V)$, $K^{M',V'}_{M'',V''} \equiv K(M',V';M'',V'')$, and
$(V_{1+2})^{M',V'}_{M'',V''} \equiv V_{1+2}(M',V';M'',V'')$.
Again, one can easily check that Equation \eqref{eq:coag2} ensures the conservation of 
the total mass density $\rho_d = \int Mf(M,V)dMdV$ for arbitrary $V_{1+2}$.

\subsection{The Volume-averaging Approximation}
In principle, the 2D Smoluchowski equation \eqref{eq:coag2} can be solved
once $K$ and $V_{1+2}$ are given.
In practice, however, direct numerical integration of Equation \eqref{eq:coag2} requires 
a huge computational expense.
To see this, let us consider that we try to solve Equation \eqref{eq:coag2} 
by dividing each parameter dimension with ${\cal N}$ fixed bins. 
At each step, we have to calculate collision events over all pairs of the bins.
For a ${\cal D}$-dimensional problems,
the number of the fixed bins is ${\cal N^{\cal D}}$, so
the total number of the collision pairs is approximately ${\cal N}^{2{\cal D}}$.
Thus, we see that a 2D calculation of the Smoluchowski equation
 is approximately ${\cal N}^2$ times heavier than an 1D one. 
Of course, one is free to take very small ${\cal N}$, but then it would result 
in a very poor resolution of $f(M,V)$. 
For a practical use, some prescription is clearly needed to manage both 
a reasonably small expense and reasonably good accuracy.
One option may be to use fewer bins and instead continuously adjust them so that they span
a parameter region with large $f(M,V)$.
This approach appears to be efficient only if the adjustment requires only a few calculations.
Another option is to adopt a direct Monte Carlo method, as done by \citetalias{OST07} and \citet{ZD08}.
Monte Carlo methods are effectively similar to the moving bin method since the Monte Carlo particles 
are continuously redistributed in the 2D space.

Now we propose a new method.
A basic strategy of this method is to give up following the details of the volume distribution and 
to only follow its {\it average} for each mass.
This enables us to calculate the porosity evolution with only $O({\cal N}^2)$ calculations at each time step as is for 1D problems.

For a rigorous formulation,
 we introduce the moments of the 2D distribution function
\beq
n(M) \equiv \int f(M,V) dV,
\eeq
\beq
\ovl{V}(M)\equiv \frac{1}{n(M)}\int V f(M,V) dV,
\eeq
where $n(M)$ and $\ovl{V}(M)$ denote the number density and {\it mean} volume of aggregates with mass $M$,
respectively.
The evolution of these quantities is described by the moment equations of the 
2D Smoluchowski equation, Equation~\eqref{eq:coag2}.
Taking the zeroth- and first-order moments of equation \eqref{eq:coag2} with respect to $V$,
we obtain 
\beqn
\frac{\pd n(M)}{\pd t} 
&=& \frac{1}{2}\int_0^M dM' \; \ovl{K}(M';M-M')n(M')n(M-M') \nonumber \\
&& - n(M)\int_0^\infty dM' \; \ovl{K}(M;M')n(M'),
\label{eq:coag20}
\eeqn
and
\beqn
\frac{\pd[\ovl{V}(M)n(M)]}{\pd t} 
&=& \frac{1}{2}\int_0^M dM' \; \ovl{V_{1+2}K}(M';M-M') \nonumber \\
&&\qquad \times n(M')n(M-M') \nonumber \\
&& - n(M)\int_0^\infty dM' \; \ovl{VK}(M;M')n(M'), 
\label{eq:coag21}
\eeqn
where $\ovl{K}$, $\ovl{V_{1+2}K}$, and $\ovl{V K}$ are the volume averages 
of $K$, $V_{1+2}K$, and $VK$ defined by
\beq
\ovl{K}(M';M'') \equiv 
\int dV'd V'' \; K^{M',V'}_{M',V''}
\frac{f_{M',V'}}{n(M')}\frac{f_{M'',V''}}{n(M'')},
\label{eq:K_avr}
\eeq
\beqn
\ovl{V_{1+2}K}(M';M'') &\equiv& 
\int dV' dV'' \; (V_{1+2})^{M',V'}_{M'',V''}K^{M',V'}_{M'',V''} \nonumber \\
&& \quad \times \frac{f_{M',V'}}{n(M')}\frac{f_{M'',V''}}{n(M'')},
\label{eq:V12K_avr}
\eeqn
\beq
\ovl{V K}(M;M') \equiv 
\int dV dV' \; V K^{M,V}_{M',V'}\frac{f_{M,V}}{n(M)}\frac{f_{M',V'}}{n(M')}.
\label{eq:VK_avr}
\eeq

Note that the above volume-averaged quantities generally depend on higher-order moments of $V$.
To solve Equations \eqref{eq:coag20} and \eqref{eq:coag21} in a closed way, 
some additional assumption about the higher-order moments is needed.

Now we try to close Equations \eqref{eq:coag20} and \eqref{eq:coag21} in terms of $n(M)$ and $\ovl{V}(M)$
by making a simple assumption.
The simplest one is that the volume distribution is so narrow
 that the volume moments higher than the first order can be ignored.
This assumption is equivalent to supposing that the distribution function $f(M,V)$ is well
approximated in the form   
\beq
f(M,V) = n(M)\delta[V-\ovl{V}(M)],
\label{eq:fdelta}
\eeq
Using this assumption, the volume integral in equations \eqref{eq:K_avr}--\eqref{eq:VK_avr} can be performed 
for arbitrary $K$ to give
\beq
\ovl{K}(M';M'') = K\bigl(M',\ovl{V}(M');M'',\ovl{V}(M'')\bigr),
\label{eq:K_avrb}
\eeq
\beq
\ovl{V_{1+2}K}(M';M'') =  V_{1+2}\bigl( M',\ovl{V}(M');M'',\ovl{V}(M'')\bigr)\ovl{K}(M';M''),
\label{eq:V12_avrb}
\eeq
and
\beq
\ovl{VK}(M;M') = \ovl{V}(M)\ovl{K}(M';M'').
\eeq
Substituting these into equations \eqref{eq:coag20} and \eqref{eq:coag21},
we obtain the closed forms of the moment equations
\beqn
\frac{\pd n(M)}{\pd t} 
&=& \frac{1}{2}\int_0^M dM' \; K^{M',\ovl{V}(M')}_{M-M',\ovl{V}(M-M')}
n(M')n(M-M') \nonumber \\
&& - n(M)\int_0^\infty dM'\;  K^{M,\ovl{V}(M)}_{M',\ovl{V}(M')}n(M'),
\label{eq:coag20b}
\eeqn
and
\beqn
\frac{\pd[ \ovl{V}(M) n(M)]}{\pd t} 
&=& \frac{1}{2}\int_0^M dM' \; 
(V_{1+2})^{M',\ovl{V}(M')}_{M-M',\ovl{V}(M-M')} \nonumber \\
&& \quad \times K^{M',\ovl{V}(M')}_{M-M',\ovl{V}(M-M')}n(M')n(M-M') \nonumber \\
&& - \ovl{V}(M)n(M)\int_0^\infty dM' \; K^{M,\ovl{V}(M)}_{M',\ovl{V}(M')}n(M'). \nonumber \\
\label{eq:coag21b}
\eeqn
In this way, we have reduced the 2D Smoluchowski equation, Equation~\eqref{eq:coag2}, 
to a closed set of two 1D equations \eqref{eq:coag20b} and \eqref{eq:coag21b} 
just by imposing a simple approximation, Equation~\eqref{eq:fdelta}. 
We refer to this assumption as {\it the volume-averaging approximation}.

The set of Equations \eqref{eq:coag20b} and \eqref{eq:coag21b} has several advantages over 
the original full 2D equation \eqref{eq:coag2}.
First, since all the volume integrals has been already performed, 
Equations \eqref{eq:coag20b} and \eqref{eq:coag21b} 
only requires ${\cal O}({\cal N}^2)$ calculations at each time step.
Second, as explained in the following subsection, 
one can solve these equations in just the same way 
as one solves the conventional 1D equation \eqref{eq:coag1}. 
The drawback is that we do not know in advance
whether the approximation made by Equation \eqref{eq:fdelta} is reasonable or not 
for a coagulation problem to be solved.
Nevertheless, we will see in Section~3 that this approximation is surprisingly  
successful in reproducing the results of previous full 2D calculations.

It is worth mentioning here that the above formulation can be extended 
to other aggregate parameters.
For example, it is straightforward to derive equations like Equations 
\eqref{eq:coag20b} and \eqref{eq:coag21b} with the electric charge $Q$ as 
an additional parameter.
This {\it charge-averaging} approximation will be particularly useful in a situation where 
aggregate collisions lead to charge separation between small and large aggregates.
Recently, \citet{Muranushi09} has shown that, with a sufficiently high dust-to-gas ratio,
the charge separation could occur for icy aggregates in protoplanetary disks.
The charge separation is potentially important since it can lead to the lightning in the disks,
which is a possible mechanism for chondrule formation \citep{DC00}.    
The charge-averaging approximation will offer a powerful tool 
for further investigation of this issue.

\subsection{Numerical Implementation}
Before proceeding to the numerical implementation of Equations \eqref{eq:coag20b} and \eqref{eq:coag21b},
we briefly review how the 1D Smoluchowski equation, Equation~\eqref{eq:coag1},
 has been solved numerically.
Let us rewrite Equation \eqref{eq:coag1} as
\beqn
\frac{\pd (Mf_M)}{\pd t} 
&=& \int_0^{M/2} dM'\; M_{1+2} K^{M'}_{M''}f_{M'}f_{M''} \nonumber \\
&& - \int_0^\infty dM'\; MK^M_{M'}f_M f_{M'},
\label{eq:coag1c}
\eeqn
where $f_M = f(M)$, $K^{M'}_{M''} = K(M';M'')$, $M'' = M-M'$, and 
$M_{1+2} = M' + M'' = M$. 
This equation means that the mass density $Mf_M$ increases 
by the collision between aggregates $M'$ and $M''(>M')$ at a rate
$M_{1+2} K^{M'}_{M''}f_{M'}f_{M''}$, 
and decreases by the collision between $M$ and $M'$ at a rate 
$M K^{M}_{M'}f_{M}f_{M'}$.
A numerical calculation of a 1D coagulation problem often
employs equation \eqref{eq:coag1c} instead of equation \eqref{eq:coag1}, 
and regards $Mf_M$ as a fundamental quantity to be evolved \citep[e.g.,][]{NNH81,THI05}.
This formulation ensures the conservation of the total mass density 
$\rho_d \equiv \int_0^\infty Mn(M)dM$ in an exact way.

Now we go back to the 2D case. 
Let us rewrite Equations \eqref{eq:coag20b} and \eqref{eq:coag21b} as 
\beqn
\frac{\pd (Mn_M)}{\pd t} 
&=& \int_0^{M/2} dM' \; M_{1+2}K^{M',\ovl{V}(M')}_{M'',\ovl{V}(M'')}
n_{M'}n_{M''} \nonumber \\
&& - \int_0^\infty dM'\; M K^{M,\ovl{V}(M)}_{M',\ovl{V}(M')}n_Mn_{M'},
\label{eq:coag20c}
\eeqn
\beqn
\frac{\pd( \ovl{V} n_M)}{\pd t} 
&=& \int_0^{M/2} dM' \; 
(V_{1+2})^{M',\ovl{V}(M')}_{M'',\ovl{V}(M'')}K^{M',\ovl{V}(M')}_{M'',\ovl{V}(M'')}n_{M''}n_{M''} \nonumber \\
&& - \int_0^\infty dM' \; \ovl{V}K^{M,\ovl{V}(M)}_{M',\ovl{V}(M')}n_Mn_{M'}, 
\label{eq:coag21c}
\eeqn
respectively, where $n_M=n(M)$.
Note that these equations are very similar to the 1D equation \eqref{eq:coag1c}.
In fact, Equation \eqref{eq:coag20c} reduces to Equation \eqref{eq:coag1c}
just by rewriting $K^{M',\ovl{V}(M')}_{M'',\ovl{V}(M'')}$ as $K^{M'}_M{M''}$. 
Furthermore, Equations \eqref{eq:coag20c} and \eqref{eq:coag21c}
are formally identical except that $M$ and $M_{1+2}$ in the former equation are 
replaced by $\ovl{V}$ and $V_{1+2}$ in the latter.
Thus, we need not to prepare any special numerical scheme 
to solve Equations \eqref{eq:coag20c} and \eqref{eq:coag21c};
we only need to adopt a scheme that has been developed to 
solve the {\it conventional} equation \eqref{eq:coag1c}.

In this study, we solve equations \eqref{eq:coag20c} and \eqref{eq:coag21c}
using a fixed-bin scheme \citep{NNH81,THI05,BDH08}. 
In this scheme, the mass space is divided into discrete bins and collision events
are calculated among the bins.
For monodisperse monomers with mass $m_\bullet$,
 the mass regions $m_\bullet \leq M \leq {\cal N}_{bd}m_\bullet$
are divided into linearly spaced bins with representative mass
 $M_k = km_\bullet \;(k =1,2,\dots, {\cal N}_{bd})$,
and the region $M > {\cal N}_{bd}m_\bullet$  are divided into 
logarithmically-spaced bins with $M_k = 10^{k/{\cal N}_{bd}}M_{k-1} (k>{\cal N}_{bd})$.
For polydisperse monomers, the mass space is just divided logarithmically.
In this study, we do not consider polydisperse monomers.
As pointed out by \citet{ONN90}, the choice of ${\cal N}_{bd}$ must be carefully done
because a small value of ${\cal N}_{bd}$ can cause artificial acceleration in coagulation
at the high-mass distribution tail. 
\citet{Lee00} performed a couple of simulations with different ${\cal N}_{bd}$ and found 
that the numerical solutions converge for ${\cal N}_{bd}>40$.
In this study, we take ${\cal N}_{bd}=40$ or $80$ (or equivalently, $M_{k+1}/M_k = 1.06$ or $1.03$)
depending on problems considered.
 
Transfer of mass and volume in the mass space are calculated in the following way.
The number $Q_{ij}$ of collision events per unit time between two bins $i$ and $j(\geq i)$
is written as $Q_{ij} = K_{ij}n_in_j$ for $j>i$ and $Q_{ij} = (1/2)K_{ii}n_i^2$ for $j=i$.
The total mass densities transferred from bins $i$ and $j$ in the events 
are $M_iQ_{ij}$ and $M_jQ_{ij}$, and the total volume densities transferred are
$\ovl{V}_iQ_{ij}$ and $\ovl{V}_jQ_{ij}$, respectively. 
As a natural consequence of the logarithmic binning, the mass of the resulting aggregates, 
$M_{ij}=M_i+M_j$, does not necessarily coincide with any of the bin masses, 
and generally goes between $M_\ell$ and $M_{\ell+1}$ with some $\ell (\geq j)$. 
Therefore, we have to determine the mass densities transferred to bins $\ell$ and $\ell+1$
in some way.
In this study, we adopt a prescription of \citet{KO69}, in which the mass densities transferred
to bins $\ell$ and $\ell+1$ are determined as $\vareps M_{ij}Q_{ij}$ and $(1-\vareps)M_{ij}Q_{ij}$
respectively, where $\vareps = (M_{\ell+1}-M_{ij})/(M_{\ell+1}-M_\ell)$.  
This prescription ensures the conservation of the total number and total mass of the resulting aggregates 
(see Appendix A.1 of \citealt*{BDH08}).
Similarly, the volume densities transferred to bins $\ell$ and $\ell+1$ 
are determined as $\vareps V_{ij}Q_{ij}$ and $(1-\vareps)V_{i+j}Q_{ij}$,
where $V_{ij} = V_{1+2}(M_i,\ovl{V}(M_i);M_j,\ovl{V}(M_j))$.
For $\ell=j$, we compute the gain and loss of the mass and volume densities 
in bin $j$ {\it simultaneously}, i.e., 
only compute the {\it net} gains $\vareps M_i Q_{ij}$ and $(V_{ij}-\ovl{V}(M_j))Q_{ij}$.
This simultaneous computation avoids systematically increasing errors in the densities
due to the near cancellation of the gain and loss terms beyond the double precision 
(see Appendix B of \citealt*{DD05}). 
Also, we adopt the ``active'' bin method \citep{Lee00,THI05}
in which collisions are not allowed for mass bins with negligibly small mass/number densities. 
This method greatly reduces the computational expense because high-mass and low-mass bins 
are mostly empty in early and later evolutionary stages, respectively.
Following \citet{THI05}, we take the critical minimum mass density 
for each mass bin to be $10^{-25}$ times the total mass density, $\rho_d$, of the system.
A test simulation by \citet{THI05} shows that 
this prescription hardly affects the evolution of the mass distribution function.

For time integration, we adopt the explicit, fourth-order Runge--Kutta method.
The time step $\Delta t$ is continuously adjusted 
so that the fractional decrease in mass density at any bin does not exceed a constant parameter $\delta$
(typically between 0.01 and 0.1).
Although we use the explicit integration scheme for simplicity,
the use of an implicit scheme \citep[e.g.,][]{BDH08} would further accelerate the calculation.

\section{Comparison between the extended Smoluchowski method and full 2D Monte Carlo methods}
To demonstrate the accuracy of our extended Smoluchowski method,
we apply this method to try to reproduce the results of full 2D Monte Carlo simulations by \citetalias{OST07}.
This problem has been also solved by \citet{ZD08}
to compare their Monte Carlo method with that of \citetalias{OST07}.

In Sections 3.1 and 3.2, we briefly summarize the collision model and the protoplanetary disk model 
assumed in the \citetalias{OST07} problem.
In Section 3.3, we show the results of our calculation and compare them with those of 
\citetalias{OST07} and \citet{ZD08}.

\subsection{The Collision Model of \citetalias{OST07}}
Here we summarize the porous aggregate model adopted by \citetalias{OST07}.
An aggregate is modeled as a cluster consisting of $N$ monodisperse dust grains (monomers)
of mass $m_0$ and radius $a_0$.  
The porosity of an aggregates is represented by the ``enlargement factor''
\beq
\psi \equiv \frac{V_A}{V_*},
\eeq
where $V_* = NV_0$ is the compact volume, and
\beq
V_A = V_0\pfrac{A}{A_0}^{3/2}
\label{eq:V_A}
\eeq
is the ``volume'' defined by the projected area $A$ of the aggregate.
$V_0$ and $A_0$ are the volume and the geometric cross section 
of a monomer given by $V_0 = (4\pi/3)a_0^3$ and $A_0 = \pi a_0^2$, respectively.
Below, we refer to $V_A$ as the area-equivalent volume.
A compact aggregate has $\psi=1$, while a porous aggregate has $\psi>1$.
Each aggregate is thus characterized by its mass $M=Nm_0$ and enlargement factor $\psi$.

The enlargement factor after a collision, 
$\psi_{1+2} = \psi_{1+2}(M_1,\psi_1;M_2,\psi_2)$, is determined in two different ways 
depending on the value of the impact energy.
The impact energy is defined by $E=(1/2)\tilde{M} (\Delta u)^2$, 
where $\tilde{M} = M_1M_2/(M_1+M_2)$ is the reduced mass and $\Delta u$ is the relative velocity.
If $E$ is smaller than the critical restructuring energy $E_{\rm restr}$,
the collided aggregates just stick to each other without compression (``hit-and-stick'' regime).
If $E>E_{\rm restr}$, the collision involves the compaction of the aggregates and another 
formula is applied for $\psi_{1+2}$ (compaction regime).
The critical restructuring energy is set to the rolling energy $E_{\rm roll}$ given by $E_{\rm roll} = (\pi a_0/2)F_{\rm roll}$, where $F_{\rm roll}$ is the rolling-friction force. 
\citetalias{OST07} adopted $F_{\rm roll} = 8.5 \times 10^{-5}(\gamma/14\,{\rm erg\,cm^{-2}})\,{\rm dyn}$
as suggested by a laboratory experiment \citep{Heim+99}.

In the hit-and-stick regime, the formula for $\psi_{1+2}$ is given by 
(Equation~(15) of \citetalias{OST07})
\beq
\psi_{1+2} = \frac{M_1\psi_1+M_2\psi_2}{M_1+M_2}\left(\frac{M_1\psi_1+M_2\psi_2}{M_1\psi_1}\right)
^{3\delta_{\rm CCA}/2-1} + \psi_{\rm add},
\label{eq:psi12_Ormel1}
\eeq
where $\delta_{\rm CCA} = 0.95$ and $\psi_{\rm add}$ is an additional factor only relevant
to BPCA-like collisions (see Equation (19) of \citetalias{OST07}). 
Rewriting equation \eqref{eq:psi12_Ormel1} in terms of $V_A$ instead of $\psi$, we obtain 
\beq
V_{A,1+2} = 
V_{A,1}\left(\dfrac{V_{A,1}+V_{A,2}}{V_{A,1}} \right)^{3\delta_{\rm CCA}/2} + \psi_{\rm add}V_{*,1+2},
\label{eq:V12_Ormel1}
\eeq
where $V_{*,1+2}=V_{*,1}+V_{*,2} = V_0(N_1+N_2)$ is the ``compact volume'' (the volume occupied by
the constituent monomers) of the newly formed aggregate.
OST07 constructed the above formula so that it agrees with the $N$-body results of \citet{Ossenkopf93}
in the BCCA and BPCA limits.
However, as we show in Section 4, the above formula is invalid between the two limits.
For $N_1+N_2 \leq 40$, the enlargement factor $\psi_{1+2}$ is directly determined by  the fitting formula of \citet{Ossenkopf93}.

In the compaction regime, $\psi_{1+2}$ is given by
\beq
V_{*,1+2}(\psi_{1+2}-1) = (1-f_C) \left[ V_{*,1}(\psi_1-1) + V_{*,2}(\psi_2-1) \right],
\eeq
where $f_C = E/[(N_1+N_2)E_{\rm roll}]$. Using the relation $V_A = V_*\psi$, we obtain the corresponding 
formula for $V_{A,1+2}$,
\beq
V_{A,1+2} = V_{A,1} + V_{A,2} - f_C(V_{A,1}+V_{A,2}-V_{*,1+2}),
\label{eq:V12_Ormel2}
\eeq
\citetalias{OST07} constructed this formula by assuming that the relative decrease in the porous volume
$V_A-V_*=V_*(\psi-1)$ scales linearly with $E$.
However, \citet{Wada+08} have recently shown that this formula does not reproduce 
the results of $N$-body simulations.

The collisional cross section $\sigma_{\rm coll}$ 
is given by $\sigma_{\rm coll} = \pi(a_{A,1} + a_{A,2})^2$, 
where 
\beq
a_A \equiv \sqrt\frac{A}{\pi} = a_0\pfrac{V_A}{V_0}^{1/3}
\label{eq:a_A}
\eeq
is the area-equivalent radius, i.e., the radius defined by the projected area $A$.

\subsection{Protoplanetary Disk Model}
\citetalias{OST07} adopted the minimum-mass solar nebula (MMSN) model \citep{Hayashi81}
in which the gas surface density $\Sigma_g$ and 
the temperature $T$ are given by power laws $\Sigma_g \propto R^{-3/2}$ 
and $T \propto R^{-1/2}$, where $R$ is the distance from the central star.
They considered two disk positions $R=1\AU$ and $5\AU$, 
and assumed rocky dust for $R=1\AU$ and icy dust for $R=5\AU$.
In this study, we consider the case of $R=1\AU$, and take the dust parameters as
$\rho_g/\rho_d = 240$, $a_\bullet = 0.1\micron$, $\rho_\bullet = 3.0~{\rm g~cm^{-3}}$, 
and $\gamma = 25~{\rm erg~cm^{-2}}$.
The strength of disk turbulence is parameterized by the so-called $\alpha$-parameter $\alpha_T$. 
\citetalias{OST07} considered three cases: $\alpha_T = 10^{-6}$, $10^{-4}$, and $10^{-2}$. 
The vertical position is taken as one scale height above the midplane.
The collision velocity is induced by Brownian motion and turbulence.
An aggregate is removed from the population (or ``rains out'' to the midplane) 
if its stopping time $\tau_f$ exceeds the critical value 
$\tau_{\rm rain} = \alpha_T/\Omega$, where $\Omega$ is the Kepler rotational frequency.
The mass of the central star is taken as $1M_\sun$, 
so that $\Omega$ is given by $\Omega = 2.0\times 10^{-7}~{\rm rad~s^{-1}}$ at $R = 1\AU$.
\citetalias{OST07} adopted $\rho_g = 8.5\times 10^{-10}~{\rm g~cm^{-3}}$, 
$c_s = 1.6 \times 10^5~{\rm cm~s^{-1}}$,\footnote{
Note that the values of $\rho_g$ and $c_s$ adopted 
by \citetalias{OST07} differ from the original MMSN values.
\citetalias{OST07} made some order of unity errors in calculating these values from $\Sigma_g$ and $T$. 
}
$\rho_g/\rho_d = 240$,
the dust material density $\rho_0 = 3.0~{\rm g~cm^{-3}}$, 
and the critical rolling energy $E_{\rm roll} = 2.4\times 10^{-9}~{\rm erg}$.

\subsection{Results and Comparison}
Following \citetalias{OST07}, we have performed simulations for four growth models: 
three porous ($\alpha_T=10^{-6},10^{-4},10^{-2}$) and
one compact ($\alpha_T=10^{-4}$).
We computed the evolution of $n(M)$ and $\ovl{V_A}(M)$ 
using Equations \eqref{eq:coag20c} and \eqref{eq:coag21c} with the numerical scheme described 
in Section 2.3.
The control parameters of the scheme are set to ${\cal N}_{bd} = 40$ and $\delta = 0.1$.
We have confirmed the numerical result is insensitive to the values of ${\cal N}_{bd}$ and $\delta$ 
as long as ${\cal N}_{bd} \geq 40$ and $\delta\leq 0.1$.
The volume $V_{A,1+2}$ of newly formed aggregates 
is calculated in accordance with the collision model of \citetalias{OST07}.

\begin{figure}
\plotone{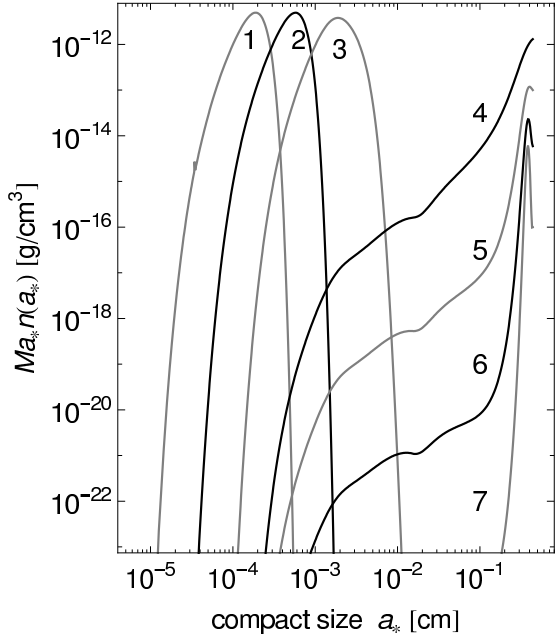}
\caption{\sf
Evolution of the mass distribution for $\alpha_T = 10^{-4}$ porous growth model
calculated using the extended Smoluchowski method described in Section 2.
The numbers 1,$\dots$,7 represent $t = 10^1{\rm yr},\dots, 10^7{\rm yr}$. 
This figure corresponds to Figure~10C in \citetalias{OST07}.
}
\label{fig:10}
\end{figure}
Figure~\ref{fig:10} shows the evolution of the mass distribution function 
for the porous, $\alpha_T = 10^{-4}$ model. 
This figure corresponds to Figure~10C in \citetalias{OST07}.
One can readily see that the evolution of our distribution function 
is qualitatively very similar to that of the \citetalias{OST07}:
nearly monodisperse growth due to Brownian motion for $t \la 10^3$ yr, 
exponential growth driven by turbulence for $10^3~{\rm yr} \la t \la 10^4~{\rm yr}$, 
and ``rain-out'' with porous size $a_A \approx 2$ cm for $t \ga 10^4~{\rm yr}$.
More detailed comparison shows that our distribution curves fall 
well inside the error bounds of their Monte Carlo results.
This means that our result agrees with that of \citetalias{OST07} even quantitatively.

\begin{figure}
\plotone{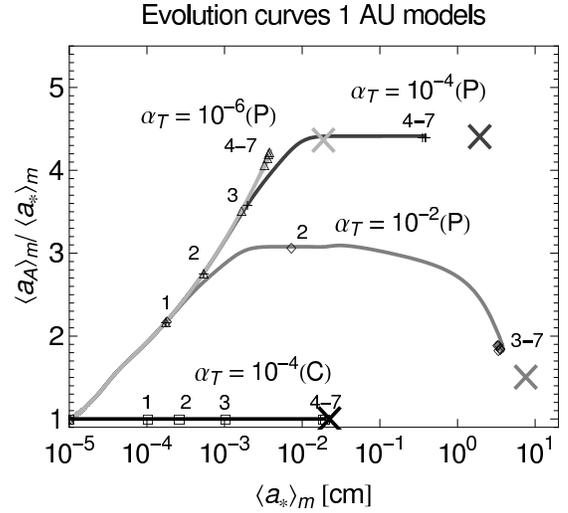}
\caption{\sf
Evolution of the mean aggregate sizes for different models,
calculated using the extended Smoluchowski method. 
$\bracket{a_A}_m$ and $\bracket{a_*}_m$ are the mass-weighted averages of the porous and compact sizes, respectively.
$\alpha_T$ denotes the turbulence parameter,
 and the letters `C' and `P' indicate the compact and porous models, respectively. 
The vertical axis $\bracket{a_A}_m/\bracket{a_*}_m$ is essentially equal to $\psi^{1/3}$.
The cross symbols indicate the average porous size $a_A$ (horizontal) 
and $a_A/a_*$ (vertical) of rain-out aggregates. 
The numbers 1,$\dots$,7 represent $t = 10^1{\rm yr},\dots, 10^7{\rm yr}$. 
This figure corresponds to Figure~12 in \citetalias{OST07}. }
\label{fig:12}
\end{figure}

Figure~\ref{fig:12} shows the evolution of the mean compact size $\bracket{a_*}$ and the mean porous size $\bracket{a_A}_m$  for different growth models.
Here, $a_* = (3V_*/4\pi)^{1/3} = N^{1/3}a_0$ is the compact (mass-equivalent) volume, 
and $\bracket{F}_m \equiv \int F(M) Mn(M)dM/\int Mn(M)dM$ is the mass-weighted average of a function $F(M)$.
The mass-weighted average approximately corresponds to the value at the peak of the mass spectrum 
$Ma(a_*)n(a_*)$. 
The vertical axis $\bracket{a_A}_m/\bracket{a_*}_m$
is roughly equal to the third root of the mean enlargement factor,
 $\bracket{\psi^{1/3}}$. 
Comparing this figure with Figure~12 in \citetalias{OST07}, we find excellent agreement 
between our evolution curves and those of \citetalias{OST07}
except that a slight difference is seen in the late-time behavior for the porous, $\alpha_T = 10^{-2}$ model. 
The cross symbols in Figure~\ref{fig:12} indicate the mean values of 
$a_A$ (horizontal axis) and $a_A/a_*$ (vertical axis) of the rain-out aggregates for the four growth models.
We compare our values with those of \citetalias{OST07} in Table~\ref{table}.
Here ``Smol+VA'' refers to the results of our extended Smoluchowski method with the volume-averaging approximation, and ``MC'' refers to the results of the full 2D Monte Carlo method 
taken from Table~2 of \citetalias{OST07}.
We find that these values agree within  errors of 10\%. 

From the above comparison, we conclude that our extended Smoluchowski method
well reproduces the results of the Monte Carlo simulations by \citetalias{OST07}.
\begin{deluxetable}{lllll}
\centering
\tablecaption{Properties of Rain-out Particles (1AU models)}
\tablecolumns{5}
\tablehead{ 
\colhead{} & \multicolumn{2}{c}{$a_A$ (cm)} & \multicolumn{2}{c}{$a_A/a_*$} \\
\colhead{Model} & \colhead{Smol+VA} & \colhead{MC} & \colhead{Smol+VA} & \colhead{MC}
 }
\startdata
$\alpha_T = 10^{-4}$, porous  & $1.9$    & $2.1 \pm 0.1$ & 4.4 & 4.4 \\
$\alpha_T = 10^{-4}$, compact & $0.022$  & $0.024$       & 1   & 1 \\
$\alpha_T = 10^{-2}$, porous  & $7.6$    & $7.8 \pm 0.7$ & 1.5 & $1.6 \pm 0.1$\\
$\alpha_T = 10^{-6}$, porous  & $0.019$  & $0.021$       & 4.4 & 4.4
\enddata
\label{table}
\end{deluxetable}

\subsection{Merits and Drawbacks of the Extended Smoluchowski Method}
In closing this section, we point out some advantages and disadvantages of our extended Smoluchowski method  over the previous Monte Carlo methods.

The most remarkable advantage of the extended Smoluchowski method is 
the efficiency in numerical calculations.
The CPU time required for our method to perform each of the above simulations is
 approximately one minute on a 3GHz CPU.
In contrast, \cite{ZD08} reported that their Monte Carlo method required 10 minutes
 for the same simulation (the calculation of \citetalias{OST07} is less efficient than this since they did not 
use a grouping algorithm as done by \citet{ZD08} and \citet{OS08}).
This means that our method is at least an order of magnitude more efficient than the Monte Carlo methods.
The high efficiency of our method is attributed to the volume-averaging approximation we introduced in Section 2.2.

Another advantage is that we can use any numerical scheme having developed for
the conventional Smoluchowski equation.
For example, one can further accelerate our method just using an implicit time integration 
developed by \citet{BDH08}. The implicit integration is particularly useful 
when one tries to include the fragmentation of aggregates 
that can make the problem extremely stiff in time \citep{BDH08,BDB09}.
By contrast, the implicit integration is inapplicable to Monte Carlo methods. 

There are two main drawbacks in our method.
First, our method cannot be used to a problem in which the volume-averaging approximation is invalid.
For example, the volume-averaging approximation will break down
 if a coagulation problem to be solved involves the fragmentation of aggregates {\it and if} the fragments 
obey broad and flat porosity distribution for each fragment mass.
For this case, we recommend improved Monte Carlo methods by \citet{OS08} and \citet{ZD08}.
Second, our Smoluchowski method is inapplicable to studying ``runaway'' 
growth in which the discreteness of the number density becomes important \citep{Wetherill90}.
Runaway growth will be well studied by the Monte Carlo code of \citet{OS08}, which
can, in principle, deal with a single aggregate. 

\section{A new hit-and-stick collision model}
The collision model of \citetalias{OST07} is only based on the knowledge of
the porosity evolution in the BCCA and BPCA limits.
This means that there are no empirical supports that validate their model 
for general types of aggregation.
In this section, we present a new collision model based on $N$-body simulations 
for more general types of aggregation.

As a first step toward a comprehensive model,
the present study focuses on ``hit-and-stick'' collisions, i.e.,
 collisions involving neither compression nor fragmentation.
The hit-and-stick approximation is valid when the collision energy $E$
is smaller than the rolling-friction energy $E_{\rm roll}$ (\citealt*{DT97}; see also Section 3.1). 
Assuming head-on collisions between equal-sized icy aggregates (made of $0.1\micron$ monomers) 
at $5\AU$ in a laminar MMSN,  for instance,
the compression and fragmentation is safely neglected until 
the monomer number of the aggregates becomes $\sim 10^{11}$,
or the porous size becomes $\sim {\rm cm}$ \citep{SWT08}.
Modeling of porosity change due to the compression and fragmentation requires $N$-body simulations 
that take into account the surface interaction between constituent particles, as recently done by
several authors \citep{Wada+07,Wada+08,Wada+09,SWT08,PD09}.
At the present, this kind of simulations have been only performed 
for limited types of collisions (e.g., equal-sized, head-on, or low-mass collisions),
and we thus have no sufficient empirical data on how the compression and fragmentation affects 
the porosity of resultant aggregates for more general cases.
For this reason, we defer the construction of the porosity change formula beyond the hit-and-stick regime
to the future work.
We also neglect the rotation of aggregates during their collision for simplicity.
The rotation could make the collisional outcome less compact \citep{KB04,PD06}.

The main goal of this section is to provide a reliable recipe for determining the porosity change
of aggregates upon general hit-and-stick collisions.
As in the previous sections, the porosity change due to a single collision are represented by 
the volume of the resultant aggregate, $V_{1+2}$.
Without loss of generality, we may rewrite this in the form
\beq
V_{1+2} = V_1 + (1+\chi)V_2,
\label{eq:V12}
\eeq
where $V_1$ and $V_2(<V_1)$ are the volumes of the collided aggregates,
 and $\chi$ is a dimensionless factor depending on the properties of the aggregates.
Since $\chi$ vanishes for the ideal compact aggregation, we can think of $\chi V_2$ as the 
volume of the ``voids'' newly created in the resultant aggregate. 
For this reason, we may regard $\chi$ as the {\it void factor}.
As we see later, the void factor $\chi$ is constant in the BCCA and BPCA limits,
suggesting that $\chi$ rather than $V_{1+2}$ is a ``good'' variable
 that describes the porosity change due to a general hit-and-stick collision. 
Our task is thus to present an empirical formula for the void factor $\chi$ in the hit-and-stick regime.

\begin{figure}
\plotone{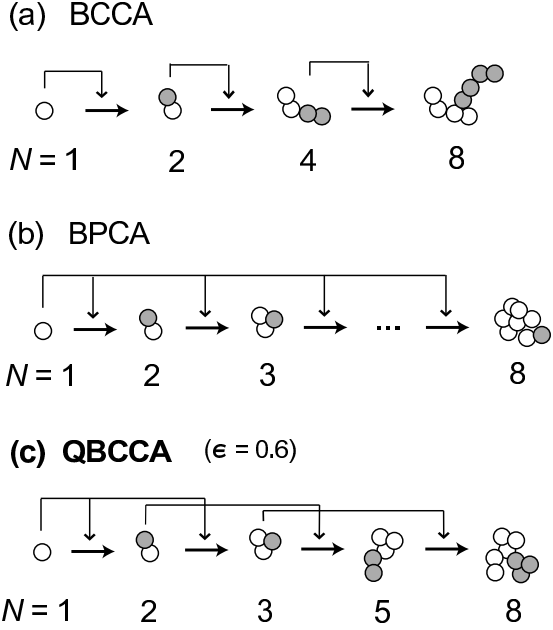}
\caption{\sf
Schematic description of various types of dust aggregation 
((a): BCCA, (b): BPCA, (c): quasi-BCCA (QBCCA) for $\eps \equiv N_2/N_1 = 0.6$).
The filled circles represent particles newly added to the aggregate at each step.
}
\label{fig:Nbody}
\end{figure}

It is essential to prepare reliable empirical data on which the porosity change formula is to be built.   
For this purpose, we have performed numerical experiments on various types of collisional aggregation.
Although classical BCCA and BPCA are useful aggregation models, these are just the two limiting cases of
general hit-and-stick collisions.
In order to bridge the gap between the two opposite limits,
we introduce a new aggregation model: {\it quasi-BCCA }(QBCCA).
We define QBCCA as a sequence of ballistic collisions between two clusters with a {\it fixed} 
mass ratio $\eps \equiv M_2/M_1 (<1)$.
Figure \ref{fig:Nbody} schematically illustrates the QBCCA model
as well as classical BCCA and BPCA. 
In the following subsections, we describe these aggregation models in more detail. 

Here we outline how we construct the collision model in this section. 
In Section 4.1, we introduce how the ``volume'' of a porous aggregate is defined in our model. 
In Section 4.2, we show the results of our $N$-body simulations to see how the volume of 
an aggregate evolves in BCCA, BPCA, and QBCCA.
In Section 4.3, we synthesize these $N$-body results to obtain a single empirical formula that 
approximately determines the value of $\chi$ for a general hit-and-stick collision.
Finally, in Section 4.4, we describe how the aerodynamical and collisional cross sections
of a porous aggregate are determined in our model.

\subsection{The Characteristic Radius and Volume}
Since a porous aggregate is generally irregular,
the concept of its ``volume,'' or ``radius,'' is somewhat ambiguous.
A useful definition for the radius is the gyration radius 
\beq
a_g \equiv \sqrt{\frac{1}{N}\sum_{k=1}^{N}(\bx_k - \bX)^2},
\eeq
where $\bx_k\,(k=1,2,\cdots,N)$ are the coordinates of constituent monomer and 
$\bX=N^{-1}\sum_k \bx_k$ is the coordinate of the center of mass.
Another definition is the {\it characteristic radius}  \citep{Mukai+92}
\beq
a_c \equiv \sqrt\frac{5}{3}a_g
= \sqrt{\frac{5}{3N}\sum_{k=1}^{N}(\bx_k - \bX)^2}.
\eeq
Using $a_c$, we can define the {\it characteristic volume}
\beq
V_c = \frac{4\pi}{3}a_c^3.
\label{eq:Vc}
\eeq
The characteristic radius has a property that it reduces to the physical radius 
for a homogeneous sphere \citep{Mukai+92}.
In our model, we regard $a_c$ and $V_c$ as the ``radius'' and ``volume'' of a porous aggregate
\footnote{For monomers ($N=1$), we simply set $a_c = a_0$ and $V_c = V_0$.}.
Note that the collision model of \citetalias{OST07} uses 
the area-equivalent volume $V_A$ (Equation~\eqref{eq:V_A}) and 
its associated radius $a_A$ (Equation~\eqref{eq:a_A}).

\begin{figure*}
\plotone{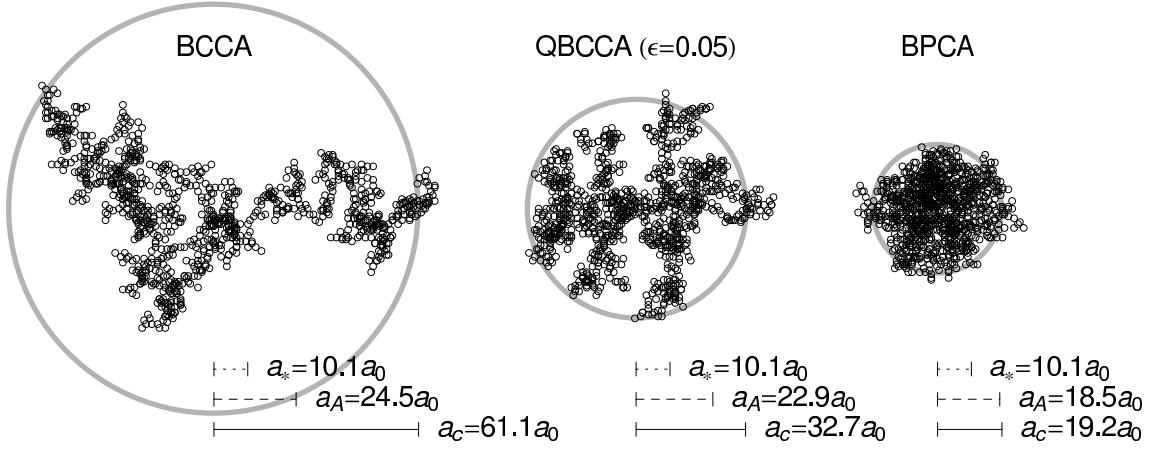}
\caption{\sf
Projection of three-dimensional BCCA (left), QBCCA ($\eps \equiv N_2/N_1 =0.048$; center),
 and BPCA (right) clusters obtained from our numerical experiments. 
Here $a_*,\, a_A,\, $ and $a_c$ denote the compact (mass-equivalent), area-equivalent,
and characteristic radii, respectively. The gray circles represent sphere of radius $a_c$
centered on the center of mass.}
\label{fig:1024}
\end{figure*}
Figure~\ref{fig:1024} shows three samples of $N$-body clusters created in our numerical experiments
together with their ``radii'' measured in different ways\footnote{
In this paper, the projected area $A$ of a cluster is calculated as the average over
15 randomly chosen orientations. 
}.
Each of the samples is composed of $\approx 10^3$ monomers, 
and hence has the compact (mass-equivalent) radius of $a_*= N^{1/3}a_0 \approx 10a_0$. 
The large circles in this figure represent the characteristic volume $V_c$ of these samples. 
We see that $a_c$ well approximates
the maximum distance from the center of mass to constituent monomers,
 but $a_A$ does not.
This fact motivates us to define the collisional cross section using $a_c$ rather than $a_A$ 
(see Section 4.4.2). 

\subsection{Porosity Evolution in Various Types of Aggregation}
Here we summarize our $N$-body experiments on three different types of aggregation 
(BCCA, BPCA, and QBCCA) and derive the volume factor $\chi$ for each of the aggregation models.

\subsubsection{BCCA}
\begin{figure*}
\plottwo{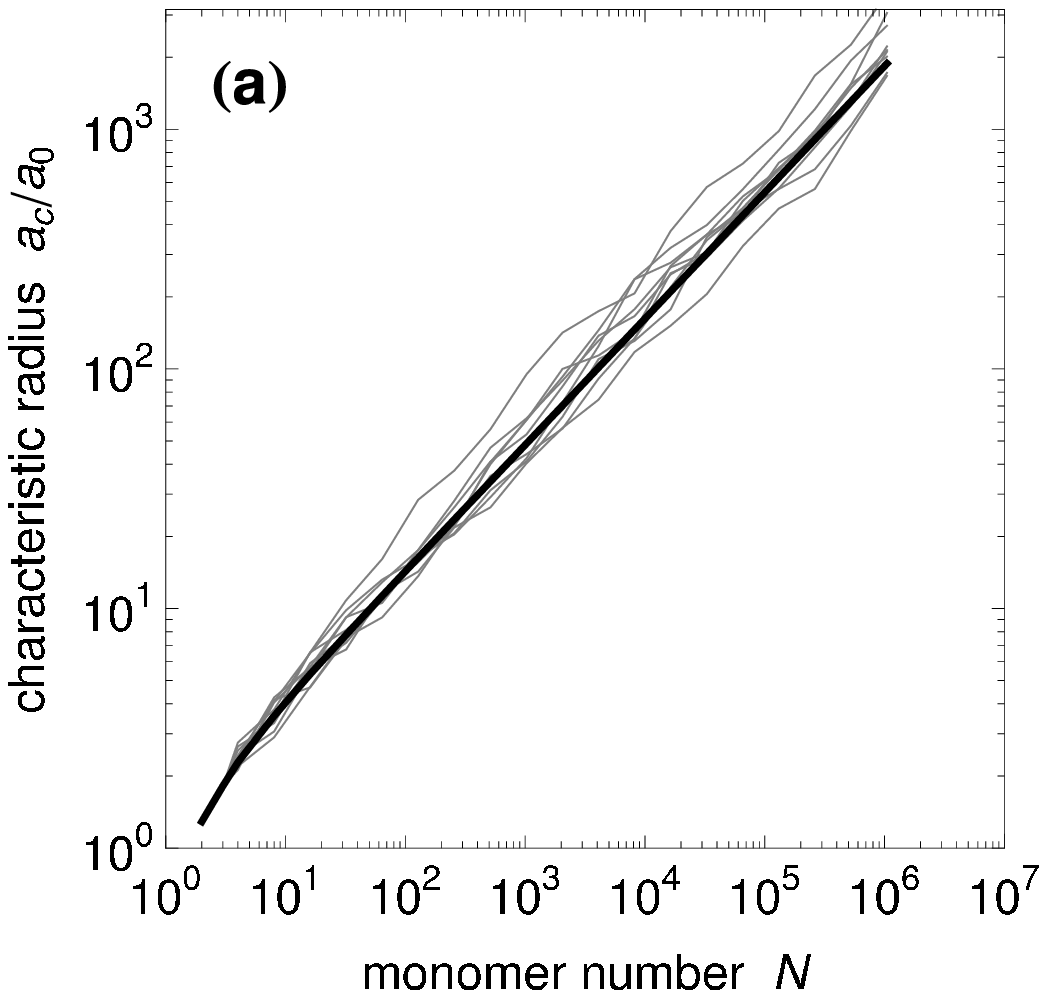}{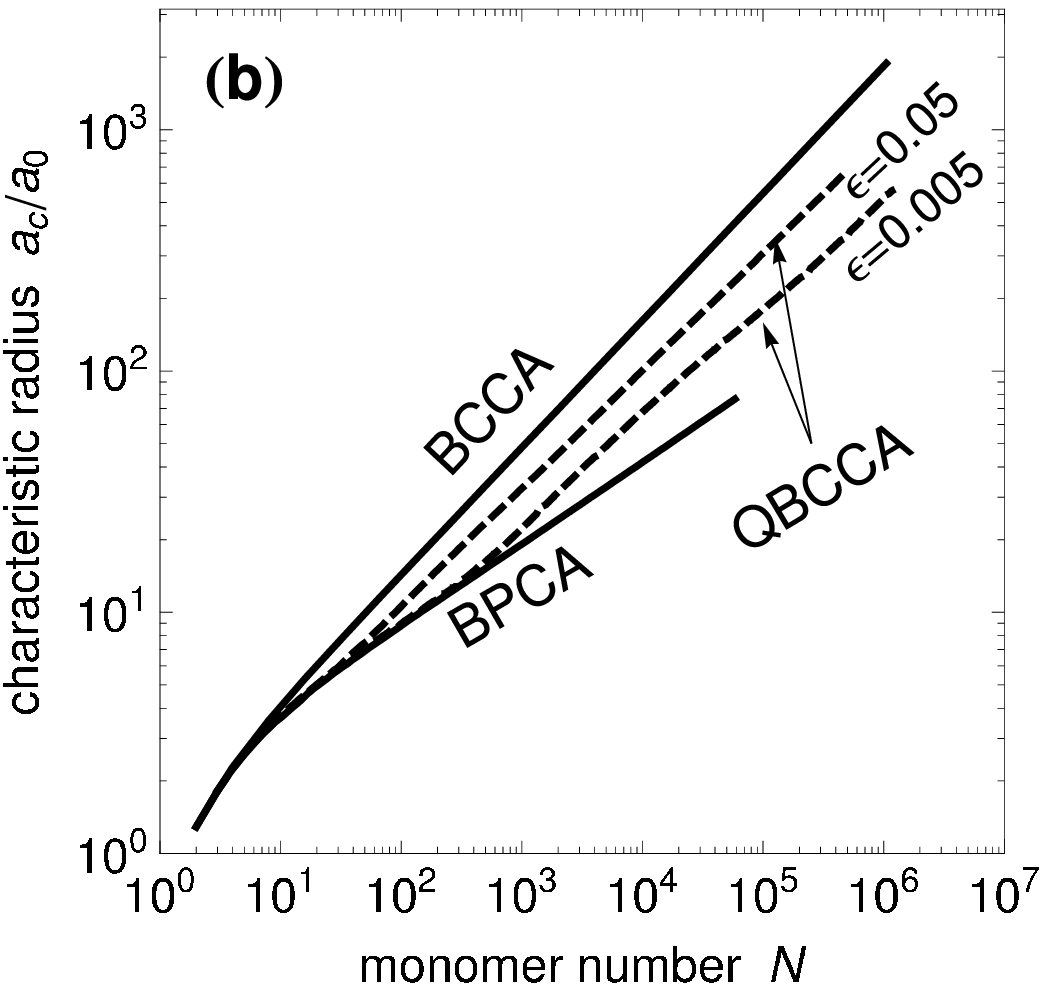}
\caption{\sf
(a) Evolution of the characteristic radius $a_c$ of BCCA clusters 
as functions of the number of constituent monomers $N$.
The gray curves indicate 10 examples of $N$-body calculation, and the average 
over $10^5$ runs is shown by the black curve.
(b)
The solid curves show the averages of $a_c(N)$ for BCCA (top) and BPCA (bottom) limits.
The averages for QBCCA models of different mass ratios $\eps = N_2/N_1$ are indicated by the dashed curves.}
\label{fig:aN}
\end{figure*}

BCCA is defined as a sequence of successive collisions between identical clusters
 (see Figure~\ref{fig:Nbody}(a)).
In the $N$-body experiments, we have simulated $10^5$ growth sequences of BCCA.
At each collisional step, we have randomly determined the impact parameter 
as well as the relative orientation of colliding aggregates, 
and followed the ballistic trajectory until one of the constituent monomers hit to another.
Figure \ref{fig:aN}{\it a} shows ten examples of the evolution of $a_c$ as a function of $N$. 
In the figure, we also show the average of $a_c$ over $10^5$ runs. 
For $N \ga 30$, the averaged $a_c$ is found to obey a single power law  
\beq
a_c \propto  N^{1/D_{\rm BCCA}},
\label{eq:aN_BCCA}
\eeq
where $D_{\rm BCCA}$ is the fractal dimension of BCCA.
Our data fitting shows $D_{\rm BCCA} \approx 1.90$,
 which is consistent with the finding of previous studies (e.g., \citealt*{Mukai+92}).

With the power-law relation \eqref{eq:aN_BCCA},
it is easy to find the void factor $\chi$ for BCCA.
Using equations \eqref{eq:Vc} and \eqref{eq:aN_BCCA}, $V_{c,1+2}$ is written as 
\beq
V_{c,1+2} = \pfrac{N_{1+2}}{N_2}^{3/D_{\rm BCCA}}V_{c,2}
\approx 2.99V_{c,2},
\label{eq:V12_BCCA}
\eeq
where $N_{1+2} = N_1+N_2 = 2N_2$.
On the other hand, Equation \eqref{eq:V12} implies $V_{c,1+2} = (2+\chi)V_{c,2}$ for BCCA. 
Comparing the two expressions, we find 
\beq
\chi \approx 0.99 \equiv  \chi_{\rm BCCA}
\label{eq:chi_BCCA}
\eeq
in the power-law limit. Note that $\chi_{\rm BCCA}$ is independent of $V_{c,1}$.

\subsubsection{BPCA}
BPCA is a sequence of successive collisions between a cluster and 
a monomer (see Figure~\ref{fig:Nbody}(b)).
In the $N$-body experiments, we have simulated $10^3$ growth sequences of BPCA 
and obtained the averaged relation between 
$a_c$ and $N$, which is plotted in Figure~\ref{fig:aN}(b).
For $N\ga 30$, the $a_c$--$N$ relation obeys a power law with a fractal dimension $\approx 3.0$,
and the porosity $P \equiv 1-V_*/V_c$ approaches to $P_{\rm BPCA} \approx 0.874$. 
This result is consistent with the finding of previous work \citep{Kozasa+93}.

Using this result, the relation between $V_c$ and $N$ is written as
\beq
V_c = \frac{V_*}{1-P_{\rm BPCA}} = \frac{V_0}{1-P_{\rm BPCA}}N.
\label{eq:Vc_BPCA}
\eeq
Since $N_{1+2} = N_1+1$, this equation implies
\beq
V_{c,1+2} = V_{c,1} + \frac{V_0}{1-P_{\rm BPCA}}.
\label{eq:V12_BPCA}
\eeq
On the other hand, equation \eqref{eq:V12} implies 
$V_{c,1+2} = V_{c,1}+ (1+\chi)V_0$ for BPCA.
Therefore we obtain
\beq
\chi = \frac{1}{1-P_{\rm BPCA}} -1 \approx 6.94 \equiv \chi_{\rm BPCA},
\label{eq:chi_BPCA}
\eeq
for the power-law limit. $\chi_{\rm BPCA}$ is independent of $V_{c,1}$ and $V_{c,2}$.
Equation \eqref{eq:chi_BPCA} is valid as long as $N\ga 30$, 
or $V_{c,1} \ga(30/1-P_{\rm BPCA})V_0 \approx 240V_0$.

\subsubsection{QBCCA}
QBCCA is defined as a sequence of successive collisions between two clusters 
with a fixed mass ratio $\eps\,(<1)$.
It is clear that BCCA corresponds to QBCCA of $\eps = 1$. 
It is also true that BPCA asymptotically approaches to QBCCA of $\eps \to 0$ 
in the limit of $N_1 \to \infty$. 
Therefore, this type of aggregation can be considered as a model between the BCCA and BPCA limits.

We here describe the general procedure of QBCCA.
Let us refer to the larger cluster as the ``target,'' and to the smaller one as the ``projectile''.
A target is always chosen to be the outcome of the latest collision.
A projectile is chosen from the outcomes of previous collisions
so that the mass ratio between the target and the projectile becomes the closest to $\eps$.
Note that this procedure is identical to that of BPCA when $N_1 \leq 1.5/\eps$, 
since the projectile is then always a monomer.
In order to obtain a ``truly'' QBCCA cluster, 
one needs to repeat the above procedure until the cluster grows beyond $N \approx 1/\eps$.   

Figure \ref{fig:Nbody} illustrates an example of QBCCA in the case of $\eps = 0.6$.
The first step is the collision between two monomers, as is for any $\eps$. 
The second collision is between the resultant dimer ($N_1=2$) and a monomer ($N_2=1$),
since $N_2=1$ is nearer to $N_1\eps = 1.2$ than $N_2=2$.
The third collision is between the resultant trimer ($N_1=3$) and a dimer $(N_2=2)$,
since $N_2=2$ is the nearest to $N_1\eps = 1.8$ among $N_2=$1, 2, and 3.
 
In the $N$-body experiments, we have chosen eight values of $\eps$ from the range $0.005\leq \eps < 1.0$.
For each value of $\eps$, we have simulated $100$--$2000$ growth sequences and obtained 
an averaged $a_c$--$N$ relation.
Figure~\ref{fig:aN}(b) shows two examples of the averaged relations ($\eps=0.05$ and $0.005$).
The averaged relations roughly obey a power low 
\beq
a_c \propto N^{1/D_{\rm QBCCA}(\eps)},
\label{eq:aN_QBCCA}
\eeq
for $N \ga 1/\eps$, where $D_{\rm QBCCA}(\eps)$ is the fractal dimension of the QBCCA clusters
 and depends on the mass ratio $\eps$.
For each value of $\eps$ we measured $D_{\rm QBCCA}$ 
by fitting the power law \eqref{eq:aN_QBCCA}
to the corresponding curve in Figure~\ref{fig:aN}(b).
Detailed inspection shows that the averaged curves for QBCCA logarithmically
oscillate around the power-law fits with a cycle $\Delta\log N \sim \log(1/\eps)$.
This is an imprint of the history that clusters of $N\la 1/\eps$ are not constructed 
by the ``truly'' QBCCA process. 
We neglect this oscillation in the following modeling since the amplitude is small.

\begin{figure}
\plotone{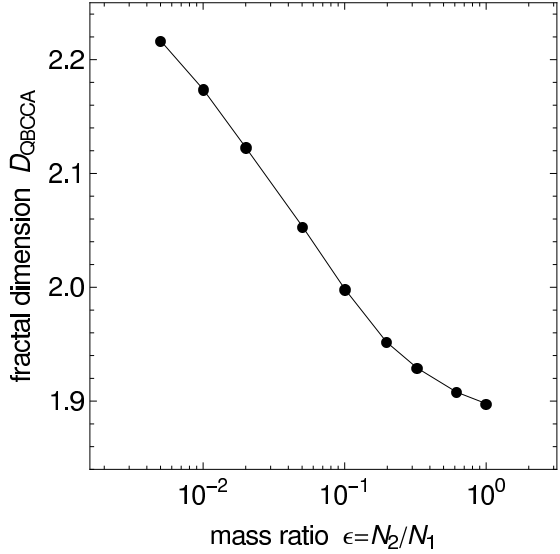}
\caption{\sf
Fractal dimension $D_{\rm QBCCA}$ of QBCCA clusters for different mass ratios 
$\eps=N_2/N_1$.
The value of $\eps=1$ corresponds to the fractal dimension $D_{\rm BCCA}$ for BCCA.
}
\label{fig:D}
\end{figure}

The obtained values of $D_{\rm QBCCA}$ as well as $D_{\rm QBCCA}(1) \equiv D_{\rm BCCA}$ 
are plotted in Figure \ref{fig:D}.
Remarkably, $D_{\rm QBCCA}(\eps)$ differs from $D_{\rm BCCA}$ by only $10\%$ 
even for $\eps \sim 10^{-2}$, or  $V_{c,2}/V_{c,1} \sim (10^{-2})^{3/D_{\rm QBCCA}} \sim 10^{-3}$.
This suggests that an aggregate growing by hit-and-stick collisions
 tends to have a fractal dimension close to 2 unless its collision partners are much smaller than 
 itself.
A simple linear extrapolation of the curve in Figure \ref{fig:D} suggests 
that $D_{\rm QBCCA}(\eps)$ will approach to 3 if $\eps$ (or $V_{c,2}/V_{c,1}$) 
is set to be as small as $10^{-7}$.
 
Equation \eqref{eq:aN_QBCCA} implies a relation similar to equation \eqref{eq:V12_BCCA},
\beq
\frac{V_{c,1+2}}{V_{c,1}} = \pfrac{N_{1+2}}{N_1}^{3/D_{\rm QBCCA}(\eps)} = (1+\eps)^{3/D_{\rm QBCCA}(\eps)},
\label{eq:V12_QBCCA}
\eeq
where we have used $N_{1+2} = (1+\eps)N_1$.
We can express this equation as a function of $V_{c,2}/V_{c,1}$ instead of $\eps$. 
First, we note that equation \eqref{eq:aN_QBCCA} also implies
\beq
\eps =  \pfrac{V_{c,2}}{V_{c,1}}^{D_{\rm QBCCA}(\eps)/3}.
\label{eq:eps}
\eeq
Since $D_{\rm QBCCA}$ is a function of $\eps$,
equation \eqref{eq:eps} determines $\eps$, and consequently $D_{\rm QBCCA}$, 
as a function of $V_{c,2}/V_{c,1}$.
Using equation \eqref{eq:eps}, equation \eqref{eq:V12_QBCCA} is written as
\beq
V_{c,1+2} = \biggl[1+\pfrac{V_{c,2}}{V_{c,1}}^{D_{\rm QBCCA}/3} \biggr]^{3/D_{\rm QBCCA}}V_{c,1}.
\label{eq:V12_QBCCA2}
\eeq
Thus, the void factor $\chi$ for QBCCA is given by
\beqn
\chi &=& \Biggl \{ 
\biggl[1+\pfrac{V_{c,2}}{V_{c,1}}^{D_{\rm QBCCA}/3} \biggr]^{3/D_{\rm QBCCA}}-1 \Biggr\}
\frac{V_{c,1}}{V_{c,2}} -1 \nonumber \\
&\equiv& \chi_{\rm QBCCA}(V_{c,2}/V_{c,1}).
\eeqn
Since $D_{\rm QBCCA}$ is a function of $V_{c,2}/V_{c,1}$, 
$\chi_{\rm QBCCA}$ only depends on $V_{c,2}/V_{c,1}$.

\begin{figure*}
\plottwo{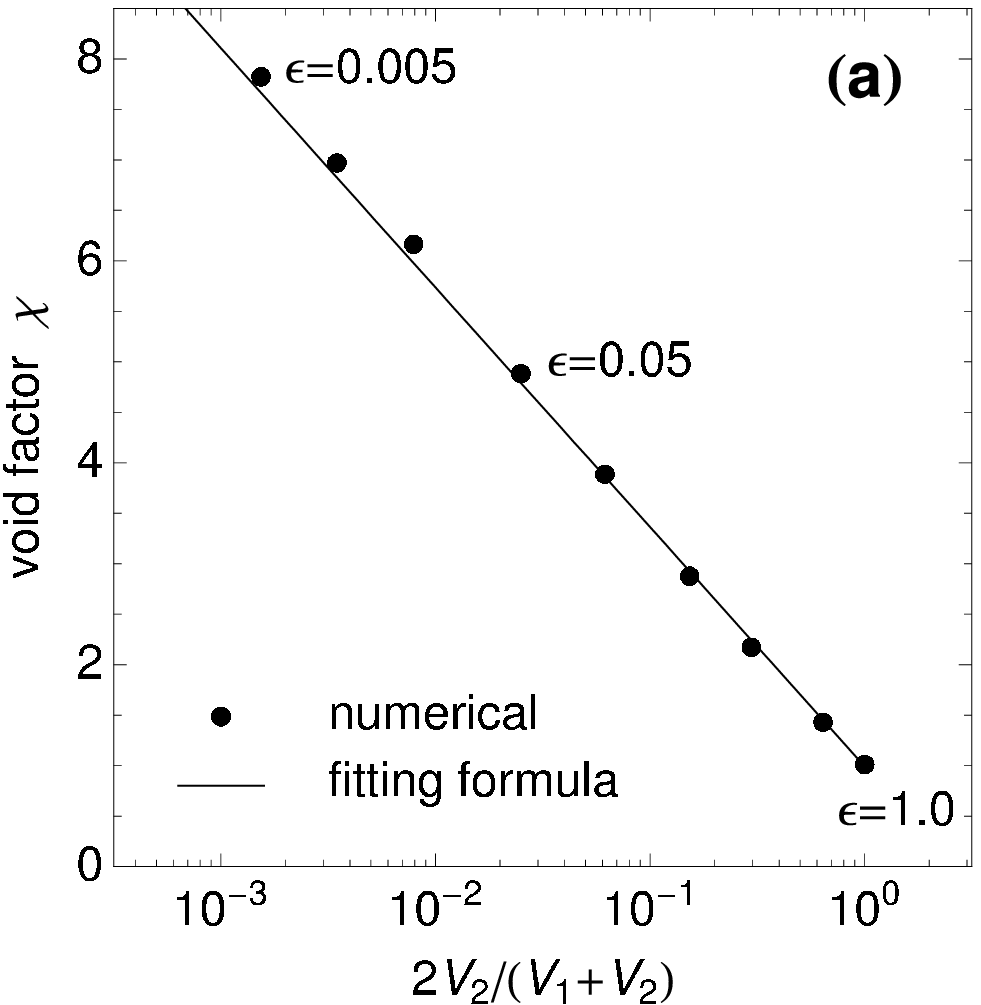}{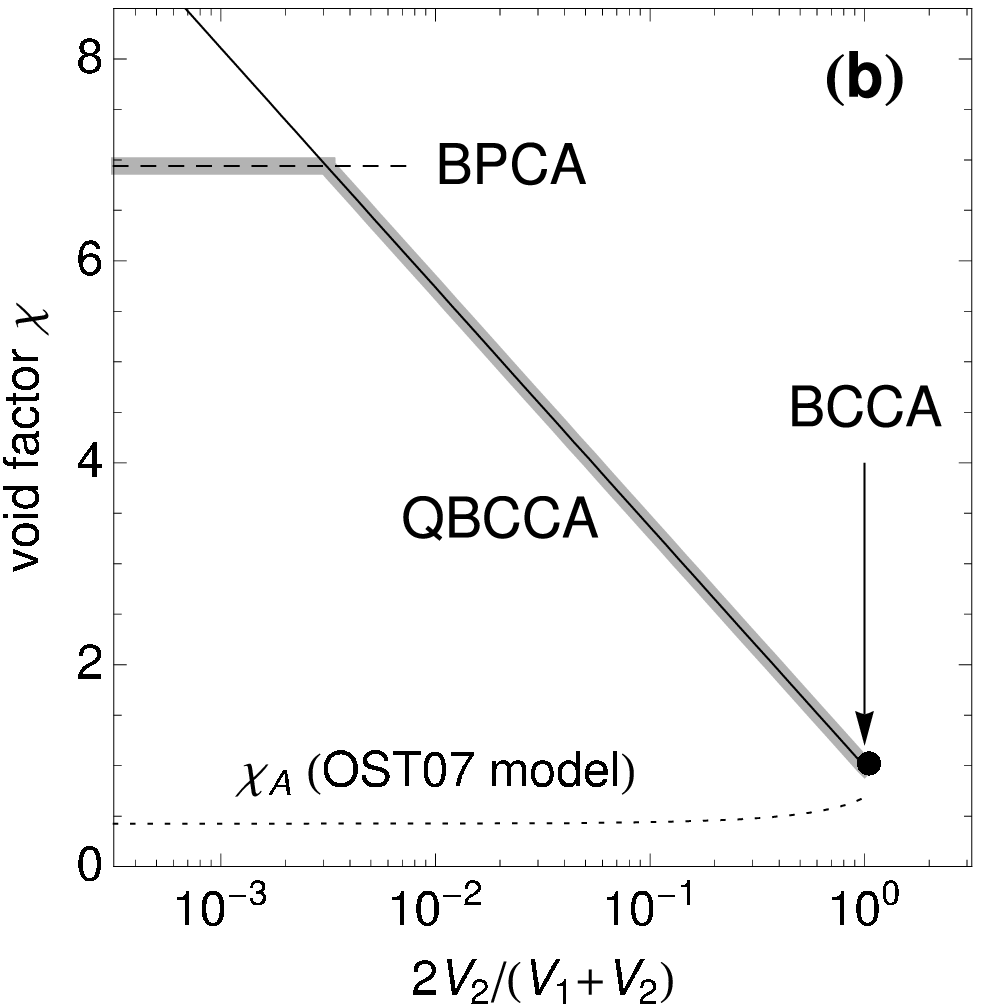}
\caption{\sf
(a) Void factor $\chi$ for QBCCA clusters as a function of $2V_{c,2}/(V_{c,1}+V_{c,2})$.
Filled circles indicate the numerical results for
various values of the mass ratio $\eps$. 
The fitting formula (Equation~\eqref{eq:chi_QBCCA}) is shown by the solid line.
(b) $\chi$ for BCCA (Equation~\eqref{eq:chi_BCCA}; filled circle), 
BPCA (Equation~\eqref{eq:chi_BPCA}; dashed line), 
and QBCCA (Equation~\eqref{eq:chi_QBCCA}; solid line).
The BPCA line is plotted  for $V_{c,2}/V_{c,1} \la 1/240$,
for which equation \eqref{eq:chi_BPCA} is valid.
The gray line indicates the void factor formula \eqref{eq:chi} for general collisions
adopted our model. 
Shown by the dotted curve is the area-equivalent void factor $\chi_A$ corresponding 
to the hit-and-stick model of \citetalias{OST07} (Equation~\eqref{eq:chi_OST07})
as a function of $2V_{A,2}/(V_{A,1}+V_{A,2})$.
}
\label{fig:chi}
\end{figure*}
Figure \ref{fig:chi} shows the void factor $\chi_{\rm QBCCA}$ as 
a function of $2V_{c.2}/(V_{c,1}+V_{c,2})= 2/(V_{c,1}/V_{c,2}+1)$.
We see that $\chi_{\rm QBCCA}$ scales linearly with $2V_{c.2}/(V_{c,1}+V_{c,2})$.
By data fitting, we obtain a simple empirical formula
\beq
\chi_{\rm QBCCA}(V_{c,2}/V_{c,1}) \approx \chi_{\rm BCCA} 
- 1.03\ln\left(\frac{2}{V_{c,1}/V_{c,2}+1}\right),
\label{eq:chi_QBCCA}
\eeq
where $\chi_{\rm BCCA}$ is given by equation \eqref{eq:chi_BCCA}.
Note that this formula includes the BCCA limit since 
it gives $\chi_{\rm QBCCA} = \chi_{\rm BCCA}$ for $V_{c,1}=V_{c,2}$.
Also note that, although $\chi_{\rm QBCCA}$ increases as $V_{c,2}$ decreases,
the void volume $\chi_{\rm QBCCA}V_2$ decreases with decreasing $V_{c,2}$.

\subsection{A Recipe for Porosity Change Due to General Hit-and-stick Collisions}
Figure \ref{fig:chi}(b) summarizes $\chi_{\rm BCCA}$, $\chi_{\rm BPCA}$, 
and $\chi_{\rm QBCCA}$ as a function of $2V_2/(V_1+V_2) = 2/(V_1/V_2+1)$.
Note that $\chi_{\rm BPCA}$ is only plotted in the range $V_{c,2}/V_{c,1}<1/240$, 
since the $a_c$--$N$ relation for BPCA obeys a power law only when $V_{c,2}/V_{c,1} \la 1/240$
(see Section~4.2.2).

Now we combine the above results to construct a recipe of the void factor $\chi$
for general hit-and-stick collisions.
In the previous subsection, we have seen that $\chi$ can be
simply written as a function of only $V_{c,2}/V_{c,1}$ under a fixed aggregation process
 (i.e., BCCA/QBCCA/BPCA).
This can be done because $N_2/N_1$ is related to  $V_{c,2}/V_{c,1}$ for each process.
This simplicity is appealing for the modeling of the porosity change. 
For this reason, we assume that {\it the growth history of an aggregate is well approximated 
either by successive collisions of a fixed volume ratio $V_{c,2}/V_{c,1}$ (BCCA/QBCCA) 
or by successive collisions of a fixed projectile volume $V_{c,2}$ (BPCA)}.
This assumption allows us to choose $\chi$ between $\chi_{\rm QBCCA}(V_{c,2}/V_{c,1})$
 and $\chi_{\rm BPCA}$. 
We will discuss the validity of this assumption in Section 5.3.
Next, we determine the choice so that the value of $\chi$ reduces to $\chi_{\rm BCCA}$ 
and $\chi_{\rm BPCA}$ in the limit of $V_{c,2}/V_{c,1} \to 1$ and $0$, respectively.  
Among the simplest ones, we  propose the following formula: 
\beq
\chi(V_{c,2}/V_{c,1}) = \min \bigl\{
\chi_{\rm QBCCA}(V_{c,2}/V_{c,1}),
\chi_{\rm BPCA}
\bigr\},
\label{eq:chi}
\eeq
where $\chi_{\rm BPCA}$ and $\chi_{\rm QBCCA}(V_{c.2}/V_{c,1})$ are given by Equations~\eqref{eq:chi_BPCA} and \eqref{eq:chi_QBCCA}, respectively.
This final formula is illustrated in Figure~\ref{fig:chi}(b)  by the thick grey line.
It should be noted here that the above choice underestimates
 the porosity increase for QBCCA with very small $V_{c,2}/V_{c,1}$.
However, as seen in Section 5.2, this effect is negligibly small. 

It is worth trying  to compare our hit-and-stick model with that of the \citetalias{OST07}.
Unfortunately, we cannot make a rigorous comparison between the two models here 
since the \citetalias{OST07} model considers the porosity increase using $V_A$.
However, it is possible to derive the ``area-equivalent void factor'' $\chi_A$
for the \citetalias{OST07} model by equating their hit-and-stick formula~\eqref{eq:V12_Ormel1} 
and $V_{A,1+2} = V_{A,1}+(1+\chi_A)V_{A,2}$.
Neglecting $\psi_{\rm add}$ in Equation~\eqref{eq:V12_Ormel1} 
(which is negligible as long as $N_1,N_2 \gg 10^2$), we obtain
\beq
\chi_{A,{\rm OST07}}(V_{A,2}/V_{A,1}) 
= \frac{V_{A,1}}{V_{A,2}}\left[\left(1+\frac{V_{A,2}}{V_{A,1}}\right)^{2\delta_{\rm CCA}/3}-1\right]-1,
\label{eq:chi_OST07}
\eeq
Note that $\chi_{A,{\rm OST07}}$ depends on $V_{A,2}/V_{A,1}$ only.
In Figure~\ref{fig:chi}(b), we overplot $\chi_{A,{\rm OST07}}$ as a function of 
 $2V_{A,2}/(V_{A,1}+V_{A,2})$.
We see that $\chi_{A{\rm OST07}}$ is approximately equal to $\chi$ of our model in the BCCA limit,
but {\it decreases} for smaller $V_{A,2}/V_{A,1}$.
As shown in Section 5, the OST07 model considerably underestimates the porosity increase
of aggregates because of this behavior. 

Summarizing Sections 4.2 and 4.3, we have constructed the porosity change recipe 
for general hit-and-stick collisions in accordance with numerical experiments on various types of aggregation.
In the numerical experiments, a new aggregation model, which we have referred as QBCCA, 
has been used to fill the gap between the conventional BCCA and BPCA models.
The porosity change recipe has been given in the form of a formula for the ``void factor'' $\chi$ (Equation~\eqref{eq:chi}),
which eventually determines the volume of a collisionally formed aggregate, $V_{1+2}$, 
via Equation~\eqref{eq:V12}.
This recipe thus allows us to evaluate the porosity change of aggregates upon a single hit-and-stick
collision.
One can calculate the growth and porosity evolution of an ensemble of dust aggregates consistently
by implementing this recipe to the Monte Carlo methods or
 the extended Smoluchowski method presented in Section 2.

\subsection{Cross Section Formulae}
The porosity of aggregates affects the aerodynamical (projected) cross section $A$ 
and the collisional cross section $\sigma_{\rm coll}$.
Below, we describe how we calculate these cross sections in our collision model.

\subsubsection{Projected (Aerodynamical) Cross Section}
The projected cross section $A$ is one of the most important properties of an aggregate since 
it determines how strongly the aggregate is coupled to gas environment.  

It is useful to see how the projected cross section behaves differently 
in BCCA and BPCA.
In the BCCA limit, $A$ is well approximated by \citep{Minato+06}
\beq
\frac{A_{\rm BCCA}}{A_\bullet}  \approx \left\{
\begin{array}{ll}
12.5N^{0.685} \exp(-2.53/N^{0.0920}), & N<16, \\[3pt]
0.352N + 0.566N^{0.862}, & N \geq 16,
\end{array} \right.
\label{eq:A_BCCA}
\eeq
where $A_\bullet=\pi a_\bullet^2$ is the geometrical cross section of a monomer.
To see how accurately this formula predicts the actual value of $A$, 
we take the BCCA cluster in Figure~\ref{fig:1024} as an example.
This cluster contains $N=1024$ monomers, and the angle average of
its projected area is $A=1890a_\bullet^2$ 
(the area-equivalent radius being $a_A = 24.5a_\bullet$). 
The empirical formula, Equation~\eqref{eq:A_BCCA}, 
gives $A_{\rm BCCA} = 1830a_\bullet^2$ ($a_A = 24.1a_\bullet$),
which agrees with the above actual value  with a relative error of $3.3\%$.
We remark here that the area-equivalent radius $a_A$ of a BCCA cluster 
is generally much smaller than its characteristic radius $a_c$.
For example, the BCCA cluster shown in Figure~\ref{fig:1024} has 
$a_c = 61.1a_\bullet$, which is about three times larger than its $a_A$.
In fact, the ``characteristic'' cross section of a BCCA cluster,
 $\pi a_c^2 \approx N^{2/D_{\rm BCCA}}A_0 \approx N^{1.05}A_0$, 
increases faster than the total geometric cross section of constituent monomers, $NA_\bullet$.
This means that if we naively assumed $A=\pi a_c^2$ in our collision model, 
we would have a BCCA cluster more and more strongly coupled to gas as its growth. 
In contrast, the formula, Equation~\eqref{eq:A_BCCA}, guarantees that  
$A$ does not increase faster than $NA_\bullet$.

In the BPCA limit, $A$ is simply related to $a_c$ as
\beq
A_{\rm BPCA} \approx \pi a_c^2,
\label{eq:A_BPCA}
\eeq
or equivalently, $a_A \approx a_c$.
For example, the BPCA cluster shown in Figure \ref{fig:1024} has 
the angle-averaged projected cross section of $A = 1080a_0^2$,
or the area-equivalent radius of $a_A = 18.5a_0$.
The characteristic radius of this cluster is $a_c = 19.2a_0$,
and thus the relative difference between $a_A$ and $a_c$ is only $4\%$.  

\begin{figure}
\plotone{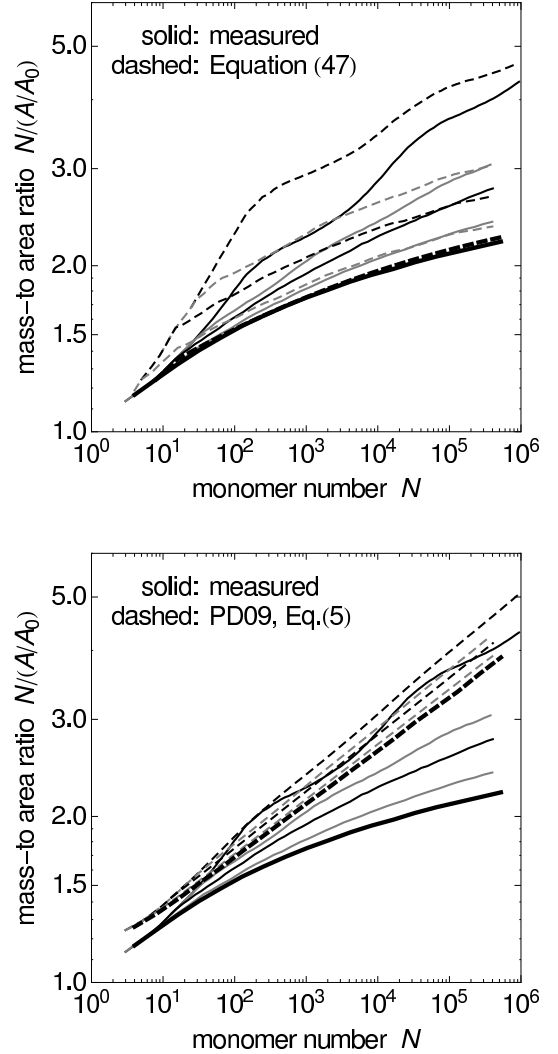}
\caption{\sf
Mass-to-area ratio $N/A$ 
of BCCA and QBCCA clusters as a function of monomer number $N$.
Upper panel: the solid curves show the directly measured values averaged over $10^2$--$10^5$ samples
for various values of $\eps$ (=1,  0.325, 0.1, 0.05, and 0.01 from bottom to top).
The oscillation seen in the curves for small $\eps$ is an imprint of the BPCA-like growth history 
for $N\la 1/\eps$ (see Section 4.2.3).
The dashed curves show the values calculated from our simple formula $A(N,a_c)$
(Equation~\eqref{eq:A}) together with the averaged $a_c$--$N$ relations. 
The good agreement between the solid and dashed curves implies that the projected cross section 
is well modeled by Equation~\eqref{eq:A} as a function of $a_c$ and $N$.
Lower panel: the solid curves are the same as those in the upper panel, but the dashed curves are obtained from Equation~(5) of \citet[Equation~\eqref{eq:A_PD09} in the present paper]{PD09}
 with $a_{\rm out} = a_c$. The formula by \citet{PD09} underestimates the real projected area
 for very large $N$.
}
\label{fig:area}
\end{figure}

Using the above facts, we construct a formula for $A$ of a general porous aggregate. 
Let us take the general formula to approximately recover Equations \eqref{eq:A_BCCA} and \eqref{eq:A_BPCA} 
in the BCCA and BPCA limits.
As one of the simplest ones, we consider
\beq
A(N,a_c) = \left( 
\frac{1}{A_{\rm BCCA}(N)} + \frac{1}{\pi a_c^2} - \frac{1}{\pi a_{c,{\rm BCCA}}^2(N)}
\right)^{-1},
\label{eq:A}
\eeq
where $A_{\rm BCCA}(N)$ is the fitting formula defined by Equation~\eqref{eq:A_BCCA} 
and $a_{c,{\rm BCCA}}(N)$ is the averaged characteristic radius of BCCA clusters
 of the same monomer number $N$.
This formula clearly reduces to Equation \eqref{eq:A_BCCA} in the  BCCA limit,
and also recovers Equations \eqref{eq:A_BPCA} in the BPCA limit for large $N$
where $a_c^2 \ll A_{\rm BCCA}$ and $a_c \ll  a_{c,{\rm BCCA}}$.
The upper panel of Figure~\ref{fig:area} shows the accuracy of this formula for more general types of clusters.
In this figure, the solid curves represent the averaged relation between the mass $N$
and the mass-to-area ratios $N/A$ obtained for various BCCA/QBCCA models
($\eps = $1,  0.325, 0.1, 0.05, and 0.01).
The dashed curves are obtained from Equation~\eqref{eq:A} 
with the averaged $a_c$--$N$ relations for our BCCA/QBCCA clusters.
For $N \la 10^6$, the relative error between the measured values and the prediction 
from Equation~\eqref{eq:A} is less than 20\% for $\eps \geq 0.05$ and 
less than 30\% even for $\eps = 0.01$.
Thus, Equation \eqref{eq:A} successfully converts $a_c$ into $A$.

Recently, \citet{PD09} have proposed another simple relation between $a_A$ and 
the outer radius $a_{\rm out} (\approx a_c)$, which reads (see Equation~(5) in \citealt*{PD09})
\beq
N\pfrac{a_0}{a_A}^3 = 1.21\pfrac{a_{\rm out}}{a_A}^{-0.3}N^{-0.33},
\label{eq:A_PD09}
\eeq
or $A \propto (N^{1.33}a_{\rm out}^{0.3})^{2/3.3} $.
They obtained this formula from relatively small ($N \la 10^3$) numerical aggregates.
The lower panel of Figure~\ref{fig:area} examines the accuracy of this formula for
larger $N$.
The solid curves are the same as those in the upper panel of Figure~\ref{fig:area},
but the dashed curves are now the prediction from Equation~\eqref{eq:A_PD09}
with $a_{\rm out}$  approximated by $a_c$.  
We see that the formula of \citet{PD09} underestimates the projected area of large ($N \ga 10^3$) 
BCCA/QBCCA clusters, especially for $\eps \ga 0.05$.

\subsubsection{Collisional Cross Section}
We explained in Section 4.1 that the characteristic radius $a_c$ of a porous aggregate 
well approximates the maximum distance from the center of mass to the constituent monomers.
It will be reasonable to regard two aggregates as collided if the impact parameter is 
smaller than the sum of their characteristic radii.
For this reason we model the collisional cross section $\sigma_{\rm coll}$ 
for two porous aggregates as
\beq
\sigma_{\rm coll} = \pi(a_{c,1} + a_{c,2})^2,
\label{eq:sigma_coll}
\eeq
where $a_{c,1}$ and $a_{c,2}$ are the characteristic radii of the aggregates.
Note that this choice differs from that of \citetalias{OST07},
 $\sigma_{\rm coll} = \pi(a_{A,1} + a_{A,2})^2$.
 
We remark here that Equation~\eqref{eq:sigma_coll} ignores the possibility that
a collision can be missed even if the impact parameter is less than $a_{c,1}+a_{c,2}$.
Such collision misses frequently occur when colliding aggregates are so fluffy 
that their fractal dimensions are much lower than two. 
For example, a microgravity experiment of \citet{KB04} implies that 
the choice of  \citetalias{OST07} agrees the actual collisional cross section better than ours
when the fractal dimension $D$ of aggregates is as low as  $1.4$.
However, the neglect of collision misses does not cause 
a serious overestimation of $\sigma_{\rm coll}$ in our model
 since we only deal with aggregates of $D >$ 1.9.

\section{Comparison of the hit-and-stick collision models}
In Section 5, we have constructed a new collision model for porous dust aggregates. 
To compare this model with the classical compact model and the model by \citetalias{OST07}, 
we apply the three models to solve simple problems on porous dust coagulation.

We consider two types of coagulation problems.
In the first type, the coagulation is purely driven by Brownian motion.
In the second type, the coagulation is driven by Brownian motion plus 
relative velocity is of the form $\Delta u = g|\tau_{f,1}-\tau_{f,2}|$,
where $g$ is a constant and $\tau_{f,1}$ and $\tau_{f,2}$ are the stopping times of two colliding aggregates.
For example, sedimentation of aggregates under a gravitational field $g$ leads to this type of relative velocity.
Another important example is the relative motion driven 
by turbulence in the strong coupling limit \citep{Weidenschilling84,OC07},
 where $g$ means the typical acceleration of the smallest turbulent eddies.
In the following, we refer to relative motion leading to the above form of $\Delta u$ 
as {\it differential drift}.

\begin{deluxetable}{lclll}
\tablecaption{Summary of hit-and-stick models}
\tablecolumns{5}
\tablehead{ 
\colhead{Model} & \colhead{$V$} &\colhead{$V_{1+2}$} 
& \colhead{$A$} & \colhead{$\sigma_{\rm coll}$} 
 }
\startdata
Compact &  $V_c$    & $V_{c,1}+V_{c,2} $ & $A_0 N^{2/3}$ & $\pi(a_{c,1}+a_{c,2})^2$ \\
OST07 &  $V_A$  & Equation~\eqref{eq:V12_Ormel1} 
& $A_0(V_A/V_0)^{2/3}$ & $\pi(a_{A,1}+a_{A,2})^2$\\
Our model & $V_c$  & Equations~\eqref{eq:V12} \& \eqref{eq:chi} 
& Equation~\eqref{eq:A} & $\pi(a_{c,1}+a_{c,2})^2$
\enddata
\label{table2}
\end{deluxetable}
We solve the above coagulation problems using the extended Smoluchowski equations
\eqref{eq:coag20b} and \eqref{eq:coag21b}.
Table~\ref{table2} summarizes the three hit-and-stick models to be compared in this section.
Here $V$ refers to the volume to be evolved by Equation~\eqref{eq:coag21b}.
Note that the characteristic volume $V_c$ is identical to  
the area-equivalent volume $V_A$ in the classical compact model.
Since the \citetalias{OST07} model gives no relation between $V_c$ and $V_A$,
we use the projected area $A$ when we compare the porosity evolution of the three models.
We set $a_{c,{\rm BCCA}}(N) = a_0N^{1/D_{\rm BCCA}}$ 
when we use Equation~\eqref{eq:A} in our model.

\subsection{Coagulation Driven by Brownian Motion}
The mean relative velocity driven by Brownian motion is written as 
\beq
\Delta u_{\rm B} = u_{\rm B\bullet}\sqrt{\frac{1}{N_1} + \frac{1}{N_2}},
\eeq
where $u_{\rm B0} = \sqrt{8k_{\rm B}T/\pi m_0}$ is the thermal velocity of a monomer.
 $\Delta u_{\rm B}$ is independent of the projected area $A$ of colliding aggregates,
 so the area formula in our model does not affect the coagulation. 
\citet{KPH99} performed a full $N$-body simulation of Brownian-motion-driven coagulation
and found that the resultant aggregates have fractal dimensions $D \approx 1.8\pm 0.2$.
We use this fact to examine the validity of our collision model.

In the numerical simulations, we started with a monodisperse ensemble of monomers and 
followed the evolution of $n(M)$ and $\ovl{V}(M)$ using Equations \eqref{eq:coag20} and \eqref{eq:coag21}. The control parameters are set to ${\cal N}_{bd} = 80$ and $\delta = 0.05$.

\begin{figure}
\plotone{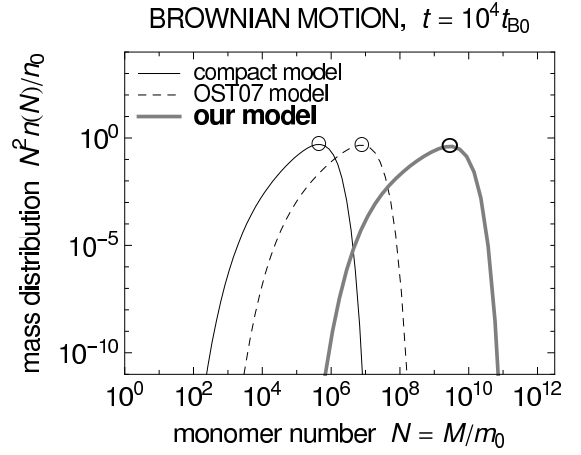}
\caption{\sf
Snapshot of the mass distribution function at $t=10^4t_{\rm B0}$ 
for Brownian-motion-driven coagulation obtained from different hit-and-stick models.
The open circles indicate the mass-weighted average masses $\bracket{M}_m$.
}
\label{fig:NdN_B}
\end{figure}

Figure \ref{fig:NdN_B} shows the normalized mass distribution functions $N^2 n(N)/n_0$ 
at time $t=10^4 t_{\rm B0}$ obtained from the three collision models.
Here $n_0$ and $t_{\rm B0}^{-1} = n_{0}\pi a_0^2 u_{\rm B0}$ are the number density 
and growth rate of monomers at the initial moment among the models.
Note that the mass conservation ensures $\int (N^2 n(N)/n_0) d\ln N  = 1$.
We find that the shape of the distribution function is insensitive to the choice of the models.
However, the growth rates of the aggregates are quite different among the models.  
At this time, the mass-weighted average mass $\bracket{M}_m$ 
is  $10^{5.7}m_0$, $10^{6.9}m_0$, and $10^{9.5}m_0$
 for the compact, \citetalias{OST07}, and our models, respectively
(see Section 3.3 for the definition of $\bracket{M}_m$).
This is essentially due to the difference in the porosity evolution among the three models
 (see Figure~\ref{fig:NA_B} below);
namely, the higher porosity results in the larger collisional cross section, and thus the
faster growth rate. 

\begin{figure}
\plotone{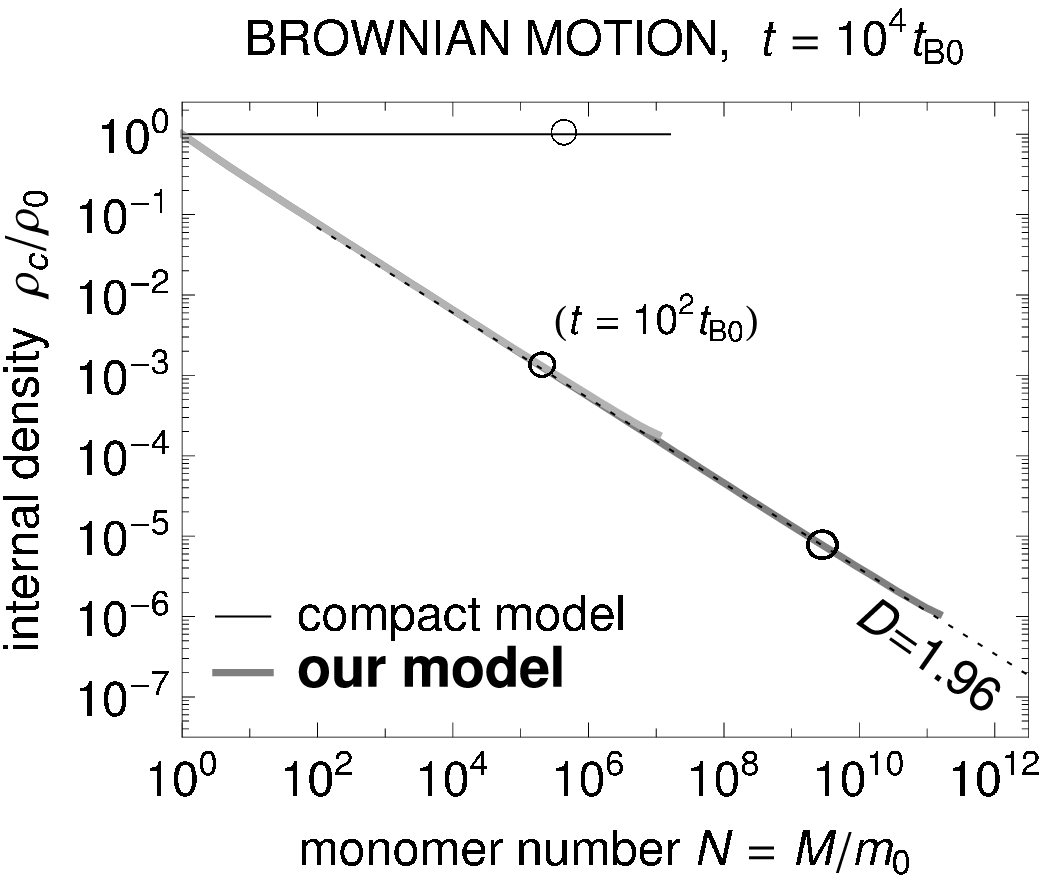}
\caption{\sf
Snapshot of the $\rho_c$--$M$ relations at $t=10^4t_{B0}$
 for Brownian-motion-driven coagulation 
obtained from the compact model and our model.
The open circles indicate the average masses $M=\bracket{M}_m$.
The dashed line shows the power-law fit to the density curve of our model, 
and $D$ is the best-fit fractal dimension.
Shown by the light gray curve is the $\rho_c$--$M$ relation for our model
 at earlier time, $t=10^2t_{B0}$.
}
\label{fig:rho_B}
\end{figure}

Figure \ref{fig:rho_B} shows the snapshots of the mean internal density
 $\rho_c(M) = M/\ovl{V}_c(M)$ at $t = 10^4t_{\rm B0}$ for the compact model and our model.
We find that the density curve for our model is well represented by 
 a single power law $\rho_c \propto M^{1-3/D}$ with $D=1.96$.
This means that the aggregates resulting from our model have a fractal dimension of $1.96$,
which is consistent with the finding of \citet{KPH99}.
Thus, our model successfully reproduces the porosity evolution of aggregates growing 
in Brownian motion. 
We also plot the density curve for our model at earlier moment, $t = 10^2t_{\rm B0}$.
The curve again lies on the same fractal line, meaning
 that the fractal dimension of the aggregates does not change with time.

The difference between the \citetalias{OST07} model and ours is best illustrated 
by the evolution of projected area $A$.
Figure \ref{fig:NA_B} shows the normalized mass-to-area ratio $N/(A/A_0)$
for the three collision models at $t = 10^4t_{\rm B0}$ as a function of $N$.
We see that the \citetalias{OST07} model gives a considerably small projected area 
for growing aggregates compared with our model.
At this moment, the aggregates of the \citetalias{OST07} model with mass $\bracket{M}_m$ 
have a projected area approximately an order of magnitude smaller than our aggregates of the same mass.
The underestimation of $A$ in the \citetalias{OST07} model is clearly caused
by the decreasing behavior of $\chi_{A,{\rm OST07}}$ for $V_{A,2}/V_{A,1}<1$.
Thus, the realistic modeling of the porosity change between the BCCA and BPCA limits is critical.
\begin{figure}
\plotone{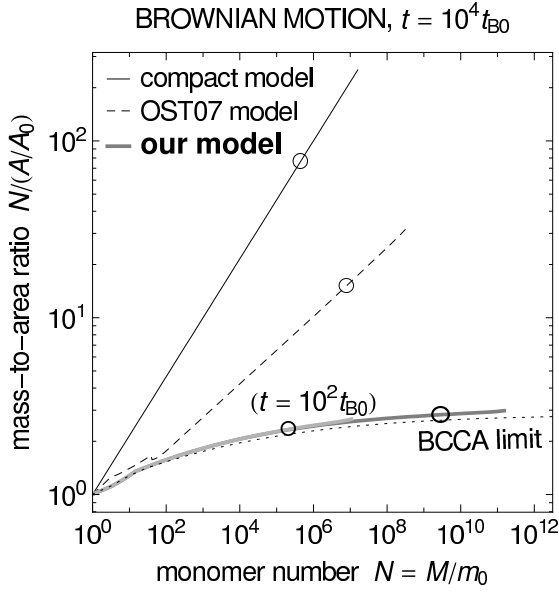}
\caption{\sf
Snapshot of the normalized mass-to-area ratios $N/(A/A_0)$ at $t=10^4t_{B0}$ 
for the Brownian-motion-driven coagulation with different collision models.
The open circles indicate the mass-weighted average masses $\bracket{M}_m$.
The  dashed curve shows the mass-to-area ratio in the BCCA limit obtained from
Equation~\eqref{eq:A_BCCA}.
Shown by the light gray curve is the $M/A$--$M$ relation for our model
 at earlier time, $t=10^2t_{B0}$.
}
\label{fig:NA_B}
\end{figure}

\subsection{Coagulation Driven by Differential Drift}
We consider aggregates smaller than the mean free path of the ambient gas,
 and adopt the Epstein's law $\tau_f \propto M/A$.
Then, the relative velocity induced by differential drift is written as 
\beq
\Delta u_{\rm D} = u_{\rm D\bullet}\left| \frac{N_1}{A_1/A_0} - \frac{N_2}{A_2/A_0}\right|,
\label{eq:v_D}
\eeq
where $u_{\rm D\bullet} = g\tau_{f\bullet}$ is the drift velocity of monomers.
\citet{WB98} conducted a laboratory experiments on dust coagulation in a turbulent gas environment 
and showed that the resultant aggregates have a fractal dimension of $D \approx 1.91$.

As in Section~5.1, we start the simulations with a monodisperse ensemble of monomers.
However, pure differential drift cannot drive the first step of the coagulation
 because $\Delta u_{\rm D}$ vanishes for all pairs of monomers,
To avoid this, we add $\Delta u_{\rm B}$ as a small perturbation 
and take the relative velocity to be 
$\Delta u = \sqrt{\Delta u_{\rm D}^2+ \Delta u_{\rm B}^2}$ with $u_{\rm B0} = 0.1 u_{\rm D0}$. 
The inclusion of $\Delta u_{\rm B}$ does not affect the dust evolution at a later time
 since $\Delta u_{\rm B}$ decreases as the aggregates grow.
The simulations are performed with the control parameters ${\cal N}_{bd} = 80$ and $\delta = 0.01$.
Time evolution is followed until $t = 20t_{\rm D0}$, 
where $t_{\rm D0} = (n_0 \pi a_0^2 u_{\rm D0})^{-1}$.

\begin{figure}
\plotone{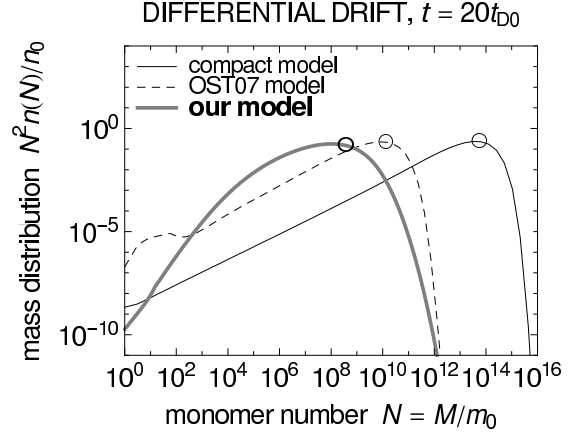}
\caption{\sf
Snapshot of the mass spectra at $t=20t_{\rm D0}$ for differential-drift-driven coagulation 
obtained with different collision models. 
The open circles indicate the positions of the mass-weighted average masses $\bracket{M}_m$.
}
\label{fig:NdN_D}
\end{figure}

Figure \ref{fig:NdN_D} compares the mass distributions $N^2n(N)/n_0$ 
for the three collision models at $t=20t_{\rm D0}$.
The average mass $\bracket{M}_m$ at this moment
is $10^{13.8}m_0$, $10^{10.1}m_0$, and $10^{8.6}m_0$ 
for the compact, \citetalias{OST07}, and our models, respectively.
In contrast to the Brownian motion case, our model results in the slowest increase
in $\bracket{M}_m$ among the three models.
This is because  our aggregates tends to conserve the mass-to-area ratio $N/A$,
and thus keep $\Delta u_D \propto \Delta(N/A)$ small, throughout the growth. 

\begin{figure}
\plotone{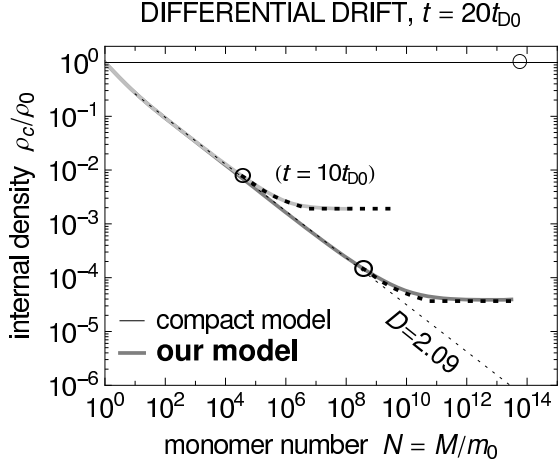}
\caption{\sf
Snapshot of the $\rho$--$M$ relations at $t=20t_{\rm D0}$ for differential-drift-driven coagulation 
obtained from the compact model and our model.
The open circles indicate the average masses $M=\bracket{M}_m$.
The dashed line shows the power-law fit to the density curve of our model, 
and $D$ is the best-fit fractal dimension.
Shown by the light gray curve is the $\rho_c$--$M$ relation for our model
 at earlier time, $t=10t_{D0}$.
The thick dotted curves show the analytical prediction by Equation~\eqref{eq:rhoratio}.
}
\label{fig:rho_D}
\end{figure}

Figure \ref{fig:rho_D} shows the snapshot of the $\rho_c$--$M$ relation at $t = 20t_{\rm D0}$
for the compact model and our model.
We see that the density curve for our model 
behaves differently between the low-mass ($M\la \bracket{M}_m$) 
and high-mass ($M \ga \bracket{M}_m$) sides.
At the low-mass side, the density curve again obeys a power-law relation.
The fractal dimension for this side is found to be $D=2.09$, which agrees with 
the finding of \citet{WB98} within an error of $10\%$.
The fractal dimension obtained here is slightly higher than that in the Brownian motion case. 
This is because the differential drift  inhibits equal-sized collisions (see Equation \eqref{eq:v_D})
and the dominant collision mode shifts to lower $\eps$.
Comparing the obtained fractal dimension $D=2.09$ with
 $D_{\rm QBCCA}(\eps)$ in Figure~\ref{fig:D},
the mass ratio of the dominant collisions is estimated as $\eps \approx 0.05$.
We discuss, in more detail, the dominant collision mode in the differential drift 
as well as Brownian motion in Section 5.3.

At the high-mass side, in contrast, the density curve deviates from 
the above power law relation and tends to flatten.
We observe this behavior independently of $t$, as is found from 
the density curve at an earlier time $(t = 10t_{D0})$ plotted in Figure~\ref{fig:rho_D}.
This implies that the growth history of massive aggregates
is qualitatively different from that of smaller ones when the coagulation is driven by the differential motion.
In fact, this high-mass behavior can be explained if we suppose that
(1) typical-mass aggregates grow by the collision with similar-sized ones, 
while (2) the massive aggregates grow by the collision with much smaller, typical-sized ($M\sim \bracket{M}_m$) ones.
For simplicity, let us divide an ensemble of aggregates into two classes, 
one being a small number of massive aggregates with mass $M_1 \gg \bracket{M}_m$,
and the other being a large population of typical aggregates with mass $M_2 \sim \bracket{M}_m$.
We ignore the collision between massive aggregates 
and assume that the both classes of aggregates grow only by colliding with typical ones.
In addition, we  assume that the both classes of aggregates grow at the same rate,
i.e.,
\beq
\frac{1}{M_1}\frac{dM_1}{dt} = \frac{1}{M_2}\frac{dM_2}{dt} 
\equiv \frac{1}{t_{\rm grow}},
\label{eq:tgrow}
\eeq
where $t^{-1}_{\rm grow}$ is the growth rate independent of $M$
\footnote{
This assumption works well when the coagulation is driven by the differential drift.
In this case, $dM/dt = \rho_d\sigma_{\rm coll}\Delta u_{\rm D}$ scales with
$(\sigma_{\rm coll}/A)M$,  so $t_{\rm grow}^{-1}$ scales with $\sigma_{\rm coll}/A$.
However, $\sigma_{\rm coll}$ is roughly proportional to $A$,
and therefore $t_{\rm grow}^{-1}$ is approximately constant. 
}.
Under these assumptions, the increase in the volume $V_2$ of a typical-mass aggregate is given by the rate of collisions with other typical-mass aggregates, $\nu_{\rm coll,22}=(dM_2/dt)/M_2 = t_{\rm grow}^{-1}$, times the volume increase per collision, 
$[1+\chi(V_2/V_2)]V_2$, i.e.,
\beqn
\frac{dV_2}{dt} &=& \frac{1+\chi_{\rm BCCA}}{t_{\rm grow}}V_2,
\label{eq:dV2dt}
\eeqn
where we have used that $\chi(V_2/V_2) = \chi_{\rm BCCA}$.
Meanwhile, the increase in the volume $V_1$ of a high-mass aggregate is given by
the rate of collisions with typical-mass aggregates,
$\nu_{\rm coll,12} = (dM_1/dt)/M_2 = (M_1/M_2)t_{\rm grow}^{-1}$, 
times the volume increase per collision, 
$[1+\chi(V_2/V_1)]V_2$, i.e.,
\beq
\frac{dV_1}{dt}
= \frac{M_1}{M_2}\frac{1+\chi(V_2/V_1)}{t_{\rm grow}}V_2 
= \frac{\rho_1}{\rho_2}\frac{1+\chi(V_2/V_1)}{t_{\rm grow}}V_1,
\label{eq:dV1dt}
\eeq
where we have used that $\rho = M/V$. Using Equations
\eqref{eq:tgrow}--\eqref{eq:dV1dt}, the time variation of $\rho_1/\rho_2$ is
written as
\beqn
\frac{d\ln(\rho_1/\rho_2)}{dt} 
&=& \frac{1}{M_1}\frac{dM_1}{dt} - \frac{1}{V_1}\frac{dV_1}{dt}
 - \left(  \frac{1}{M_2}\frac{dM_2}{dt} - \frac{1}{V_2}\frac{dV_2}{dt} \right)
 \nonumber \\
&=& - \frac{1}{V_1}\frac{dV_1}{dt} + \frac{1}{V_2}\frac{dV_2}{dt} \nonumber \\
&=& \frac{1}{t_{\rm grow}} 
\left\{ 1+ \chi_{\rm BCCA} 
- \frac{\rho_1}{\rho_2}\left[ 1+\chi(V_2/V_1)\right] \right\}.
\eeqn
The solution to this equation asymptotically satisfies $d(\rho_1/\rho_2)/dt = 0$, or
\beq
\rho_1
= \frac{1+\chi_{\rm BCCA}}{1+\chi(V_2/V_1)}\rho_2 \nonumber \\
= \dfrac{1+\chi_{\rm BCCA}}{1+\chi\left(\frac{M_2/M_1}{\rho_2/\rho_1}\right)}\rho_2.
\label{eq:rhoratio}
\eeq
For fixed $M_2$ and $\rho_2=\rho(M_2)$, 
this final equation implicitly determines $\rho_1$ as a function of $M_1$. 
In Figure \ref{fig:rho_D}, we overplot the solutions $\rho_1$ to Equation \eqref{eq:rhoratio} 
at $t = 10t_{D0}$ and $20t_{D0}$ as a function of $M_1$ with $M_2 = \bracket{M}_m$ and $\rho_2=\rho(\bracket{M}_m)$.
We find that these solutions reproduce the flattening of the density profiles.
The flattening of the density curve is intuitively reasonable, since 
 the porosity of an aggregate is generally kept small when it only accretes much smaller aggregates.

\begin{figure}
\plotone{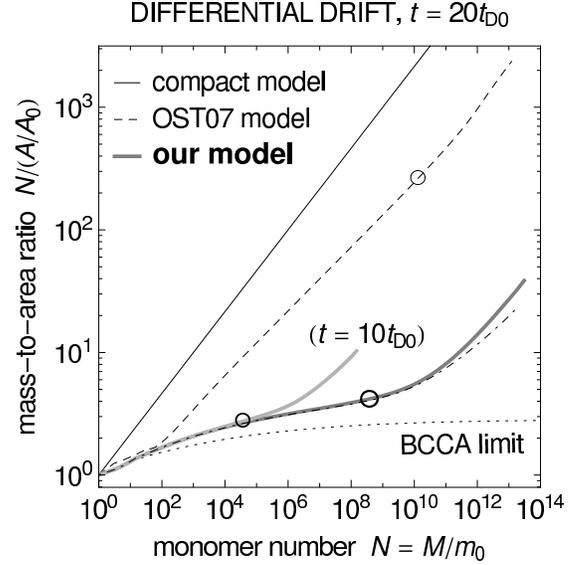}
\caption{\sf
Snapshot of the normalized mass-to-area ratios $N/(A/A_0)$ at $t=20t_{D0}$ 
for the differential-drift-driven coagulation with different collision models.
The open circles indicate the mass-weighted average masses $\bracket{M}_m$.
The  dashed curve shows the mass-to-area ratio in the BCCA limit obtained from Equation~\eqref{eq:A_BCCA}.
The dot-dashed curve is the result obtained from our model with setting $\chi = \chi_{\rm QBCCA}$ 
for all $V_2/V_1$ instead of Equation~\eqref{eq:chi}.
Shown by the light gray curve is the $M/A$--$M$ relation for our model
 at earlier time, $t=10t_{D0}$.
}
\label{fig:NA_D}
\end{figure}
Figure~\ref{fig:NA_D} compares the normalized mass-to-area ratios $N/(A/A_0)$ for the two collision models  at $t = 20t_{\rm D0}$ as a function of $M$. 
The mass-to-area ratio is a key property of aggregates growing by differential drift
since the terminal velocity is proportional to it.
Again, the $M/A$--$M$ relation for our model exhibits different behavior 
between the low-mass $(M\la \bracket{M}_m)$ and high-mass $(M\ga \bracket{M}_m)$ sides:
nearly flat in the low-mass side and increasing with mass in the high-mass side. 
This difference clearly reflects the $\rho$--$M$ relation seen in Figure~\ref{fig:rho_D}.
Moreover, detailed inspection shows that $M/A$ for the \citetalias{OST07} model 
 also steepens in the high-mass end.
Thus, the steepening of the $M/A$ curve is not peculiar to our model. 

One might suspect that the increase in the mass-to-area ratio
is caused by the cutoff in the void factor $\chi$ (Equation~\eqref{eq:chi}) at small $V_2/V_1$.
Actually,  the complete flattening as seen in Figure~\ref{fig:rho_D} is achieved because 
we have chosen $\chi = \chi_{\rm BPCA}$ in the small $V_2/V_1$ limit (see Section 4.3).
This can be confirmed from Equation~\eqref{eq:rhoratio}, which implies
$\rho_1 \to  \rho_2 (1+\chi_{\rm BCCA})/(1+\chi_{\rm BPCA})\approx 0.25\rho_2$
independently of $V_1$.
However, flattening still occurs even if we choose $\chi$ 
 to be $\chi_{\rm QBPCA}(V_2/V_1)$ for all $V_2/V_1$.
In fact, Equation~\eqref{eq:rhoratio} then implies 
$\rho_1 \propto 1/\log(V_1/V_2)$ in the small $V_2/V_1$ limit,
 which decreases only logarithmically with increasing $V_1$. 
For confirmation, we have solved the same coagulation problem by 
removing the cutoff in $\chi$ and assuming $\chi = \chi_{\rm QBCCA}(V_2/V_1)$ for all $V_2/V_1$.
The result is overplotted in Figure~\ref{fig:NA_D}.
We see that $M/A$ at the high-mass end is mostly unchanged even if the cutoff in $\chi$ 
is removed.
Thus, the increase of the mass-to-area ratio seems to be a robust feature 
for the coagulation driven by the differential drift.
  
The above finding has a potentially important implication for
 the growth of dust aggregates in protoplanetary disks.  
One of the authors has recently examined the effect of 
 dust charging on its collisional growth in protoplanetary disks \citep{Okuzumi09}.
Aggregates in a disk usually charge up negatively on average, and their collisions
require a sufficiently high collision velocity to overcome the electrostatic barrier.
Strikingly, it turned out that the electrostatic repulsion strongly suppresses
the growth of fractal ($D \approx 2$) aggregates.
This is essentially because a fractal aggregate of $D \approx 2$ keeps
 a small mass-to-area ratio and thus a low sedimentation velocity.
However, our calculation shows that high-mass aggregates in an ensemble 
have a large mass-to-area ratio and a high sedimentation velocity compared with 
low-mass fractal aggregates.
This implies that the high-mass aggregates may be allowed to continue growing 
even if the others are not allowed.
In a forthcoming paper, we will investigate this issue 
using the numerical tools presented in this study.

\subsection{Validity and Possible Limitation of Our Porosity Evolution Formula}
In the modeling of the porosity evolution, we have assumed that 
the growth of an aggregates is well approximated as BCCA/QBCCA or BPCA.
Here, we examine whether this assumption holds in the above coagulation problems. 

\begin{figure}
\plotone{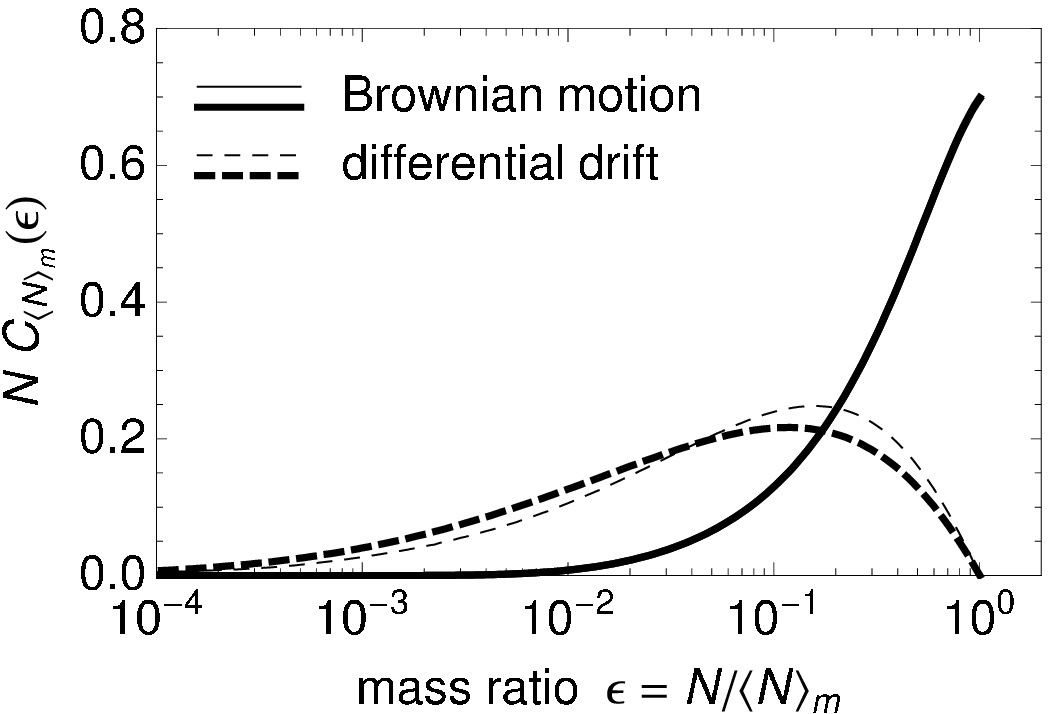}
\caption{\sf
Normalized, mass-weighted collision rates $C_{\bracket{N}_m}(N)$ 
between aggregates of masses $\bracket{N}_m$ and $N=\eps \bracket{N}_m$ 
as a function of $\eps = N/\bracket{N}_m$ 
(thick solid: Brownian motion, $t = 10^4t_{B0}$; thin solid: Brownian motion, $t = 10^2t_{B0}$;
thick dashed: differential drift, $t = 20t_{D0}$; thin dashed: differential drift, $t = 10t_{D0}$).
Note that the two solid curves are indistinguishable. 
}
\label{fig:coll}
\end{figure}

Let us consider a quantity
\beq
C_{N_1}(\eps) \equiv 
\frac{\eps \ovl{K}(\eps N_1;N_1)n(\eps N_1)}{\dint_0 ^{1}\eps' \ovl{K}(\eps'N_1;N_1)n(\eps' N_1)d\eps'},
\eeq
where $\ovl{K}(\eps N_1;N_1)n(\eps N_1)$ is the rate of collisions
between aggregates with mass $\eps N_1$ and $N_1$.
The growth rate $t_{\rm grow}^{-1} = ({dN_1}/{dt})/N_1$
of an aggregate with mass $N_1$ is written as 
$t_{\rm grow}^{-1} = {\int_0 ^{1}\eps' \ovl{K}(\eps'N_1;N_1)n(\eps' N_1)d(\eps' N_1)}$.
Thus, $C_{N_1}(\eps)d\eps$ measures how collisions of mass ratios between
$\eps$ and $\eps+d\eps$ contribute to the growth of an aggregate of mass $N_1(\geq \eps N_1)$.
Figure~\ref{fig:coll} plots $C_{\bracket{N}_m}(\eps)$ for Brownian motion and differential drift
at different times $t$.
For both the cases, the functional form of $C_{\bracket{N}_m}(\eps)$
is nearly independent of $t$ and has a peak at $\eps = 1$ (Brownian motion case) 
and $\eps \sim 0.1$ (differential drift case).  
The distribution of  $C_{\bracket{N}_m}(\eps)$ has  an order-of-magnitude spread
 in $\eps$-space.
However, this spread can be regarded as narrow since an order-of-magnitude fluctuation 
in $\eps$ only causes $\sim 10\%$ fluctuation in $D_{\rm QBCCA}$ (see Figure~\ref{fig:D}).
Therefore, the growth of aggregates is indeed well approximated as BCCA/QBCCA
 when the coagulation is driven by Brownian motion and differential drift.
  
On the other hand, the assumption may become less good if the collision partners of an aggregate 
have widely spread mass ratios and if they contribute equally to its growth.
Such a situation may be realized when, for example, the collision partners obey a very broad 
and flat size distribution.    

\section{Summary}
In this study, we have presented new numerical tools for studying the coagulation 
and porosity evolution of dust aggregates. Our findings are summarized as follows.

\begin{enumerate}
\setlength{\itemsep}{-0pt}
\item
We have presented a new numerical method for simulating the coagulation and porosity evolution 
of dust aggregates as an extension of the conventional Smoluchowski method (Section 2). 
The new method treats the averaged volume of aggregates of the same mass
as dynamical,
and follows its evolution as well as the evolution of the mass distribution function consistently
(Equations \eqref{eq:coag20b} and \eqref{eq:coag21b}).
This method enables to treat the 2D (i.e., mass and porosity being dynamical)
 coagulation problems in a very efficient way.
We have confirmed that our method well reproduces the results of previous 
full 2D Monte Carlo simulations with much fewer computational expense (Section~3).

\item
We have presented a new collision model based on our numerical experiments
 on aggregates collisions (Section~4).
A collision model refers to a set of definitions and formulae for the properties of porous aggregates, 
including a recipe for the porosity change upon a collision.
As the first step, we have focused on ``hit-and-stick'' collisions,
i.e., collisions involving no restructuring or fragmentation.
In the numerical experiments, we have first considered successive collisions between 
aggregates of a fixed mass ratio, which we have referred as QBCCA.
This allows to fill the gap between the classical BCCA and BPCA.
Using the results of the $N$-body experiments on BCCA, BPCA, and QBCCA, 
we have constructed a recipe for the porosity change due to a general hit-and-stick collision
 (Equation~\eqref{eq:chi}) as well as formulae for the aerodynamical and collisional cross sections.
In a forthcoming paper, we will include the effects of collisional compression and fragmentation
into our collision model.

\item
To clarify the validity of our collision model and the difference from previous models,
we have performed a couple of simple simulations on porous dust coagulation
with the extended Smoluchowski method (Section 5).
We considered two cases where the coagulation is driven by Brownian motion 
and differential drift, respectively.
For both cases the simulations using our model result in fractal aggregates of $D\approx2$,
which is consistent with the findings of 
previous full $N$-body and laboratory experiments \citep{KPH99,WB98}.
By contrast, the simulations using the hit-and-stick model of \citetalias{OST07} 
result in much less porous aggregates.
This is because this model underestimates the porosity increase upon unequal-sized collisions. 

\item
We have discovered that, when the coagulation is driven by differential drift,
aggregates at the high-mass end of the mass distribution 
obey a flat density--mass relation, in marked contrast to aggregates 
of lower masses, which obey a fractal (i.e., power-law) density--mass relation (Section 5.2).
This is explained by the difference in the dominant growth modes:
typical aggregates grow by colliding with similar ones,
while the most massive aggregates grow by colliding with much smaller ones.
As a consequence of this tendency, the massive aggregates can have a
 drift velocity ($\propto$ mass-to-area ratio) considerably higher than smaller ones.
This finding may be crucially important in protoplanetary disks where
the growth of slowly moving aggregates is suppressed due to the negative charging.
\end{enumerate}
  
\acknowledgments
We thank Andras Zsom for providing us with 
useful information on his Monte Carlo simulations.
We also thank the referee, Chris Ormel,  for fruitful comments leading
 to significant improvements in the manuscript.

\end{document}